\documentclass[british,10pt,final,twocolumn,twoside]{IEEEtran}
\usepackage[T1]{fontenc}
\usepackage[utf8]{inputenc}
\usepackage{xcolor}
\usepackage{pdfcolmk}
\usepackage{verbatim}
\usepackage{amsmath}
\usepackage{amssymb}
\usepackage{graphicx}
\usepackage{esint}
\PassOptionsToPackage{normalem}{ulem}
\usepackage{ulem}

\makeatletter

\providecolor{lyxadded}{rgb}{0,0,1}
\providecolor{lyxdeleted}{rgb}{1,0,0}

\newenvironment{lyxlist}[1]
{\begin{list}{}
{\settowidth{\labelwidth}{#1}
 \setlength{\leftmargin}{\labelwidth}
 \addtolength{\leftmargin}{\labelsep}
 }}
{\end{list}}

\usepackage{graphicx}
\usepackage{cite}       

\global\long\def\diag{\operatorname{diag}}
\global\long\def\trace{\operatorname{tr}}

\global\long\def\erfc{\operatorname{erfc}}

\global\long\def\meter{\,\text{m}}

\newcommand{\switchAtwo}[2]{#1}
\newcommand{\switchBtwo}[2]{#1}

\newcounter{MYtempeqncnt}

\usepackage[british]{babel}

\makeatother

\usepackage{babel}
\begin{document}

\title{On the Performance Limits of Map-Aware Localization}

\author{Francesco Montorsi,~\IEEEmembership{Student Member,~IEEE}, Santiago
Mazuelas,~\IEEEmembership{Member,~IEEE}, \\
Giorgio M. Vitetta,~\IEEEmembership{Senior Member,~IEEE}, Moe Z.
Win,~\IEEEmembership{Fellow,~IEEE}%
\thanks{This research was supported, in part, by the Air Force Office of Scientific
Research under Grant FA9550-12-0287, the Office of Naval Research
under Grant N00014-11-1-0397, and the MIT Institute for Soldier Nanotechnologies.%
}%
\thanks{Francesco Montorsi and Giorgio M. Vitetta are with the Dept. of Information
Engineering, University of Modena and Reggio Emilia, Modena, Italy
(e-mail: francesco.montorsi@unimore.it; giorgio.vitetta@unimore.it).%
}%
\thanks{Santiago Mazuelas and Moe Z. Win are with the Laboratory for Information
and Decision Systems (LIDS), Massachusetts Institute of Technology,
Cambridge, MA 02139 USA (e-mail: mazuelas@mit.edu; moewin@mit.edu).%
}%
}
\maketitle
\begin{abstract}
Establishing bounds on the accuracy achievable by localization techniques
represents a fundamental technical issue. Bounds on localization accuracy
have been derived for cases in which the position of an agent is estimated
on the basis of a set of observations and, possibly, of some a priori
information related to them (e.g., information about anchor positions
and properties of the communication channel). In this manuscript new
bounds are derived under the assumption that the localization system
is \emph{map-aware}, i.e., it can benefit not only from the availability
of observations, but also from the a priori knowledge provided by
the \emph{map} of the environment where it operates. Our results show
that: a) map-aware estimation accuracy can be related to some features
of the map (e.g., its shape and area) even though, in general, the
relation is complicated; b) maps are really useful in the presence
of some combination of low signal-to-noise ratios and specific geometrical
features of the map (e.g., the size of obstructions); c) in most cases,
there is no need of refined maps since additional details do not improve
estimation accuracy.\end{abstract}

\section{Introduction\label{sec:intro}}

\noindent Recently, localization systems have found
widespread application, since they allow to develop a number of new
services in both outdoor and indoor environments \cite{Pahlavan2002,Win2011}.
Conventional localization systems acquire positional information from
a set of observations; these are usually extracted from noisy wireless
signals propagating in harsh environments. Unavoidably, this limits
the accuracy that can be achieved by such systems.

\noindent Localization accuracy can certainly benefit from the availability
of any form of \emph{a priori} knowledge. In the technical literature,
the sources of prior knowledge commonly exploited are represented
by the positions of some specific nodes (called \emph{anchors}) or
by some characteristic of the communication channel; for instance
in \cite{Qi2002,Larsson2004,Gustafsson2005,Qi2006} the impact of
different factors (like non-line-of-sight, NLOS, propagation and network
synchronization) on localization accuracy have been thoroughly analysed
employing Cramer-Rao bounds (CRBs) and, when a priori knowledge is
available, Bayesian Cramer-Rao bounds (BCRBs). In \cite{Gezici2005,Gezici2009,Shen2010,Shen2010a}
multipath propagation in ultra wide band (UWB) systems and its effects
on time-of-arrival (TOA) estimation have been investigated by means
of CRBs or BCRBs. In \cite{Patwari2005,Shen2010b,Mazuelas2011} CRBs
and BCRBs for the analysis of cooperative localization techniques
have been derived.

\noindent In recent times, however, some attention has been paid to
the possibility of improving accuracy by endowing localization systems
with \emph{map-awareness}, i.e., with the knowledge of the \emph{map}
of the environment in which they operate. Specific examples of map-aware
algorithms have been developed in \cite{Singh2007,Klepal2008,Davidson2010,Anzai2010,Anisetti2011},
which propose the adoption of non-linear filtering techniques (namely,
particle filtering and extended Kalman filtering) to embed map-based
a priori information in navigation systems. This work evidences that
this kind of information plays an important role in improving localization
and navigation accuracy; however, the impact of map-awareness on the
performance limits of localization systems is still an open problem.
In fact, as far as we know, the technical literature dealing with
the fundamental limits of localization accuracy \cite{Shen2010a,Shen2010b,Shen2010,Mazuelas2011}
has not considered this issue yet. 

\noindent In this paper, novel accuracy bounds for map-aware localization
systems are developed and their applications to specific environments
are analysed. Specifically, the BCRB, the extended Zik-Zakai bound
(EZZB) and the Weiss-Weinstein bound (WWB) for the above mentioned
systems are derived. These bounds, which provide several novel insights
into map-aware localization, have the following features: 
\begin{enumerate}
\item they are characterized by different tightness/analytical complexity;
\item \noindent they can be evaluated for any map geometry and, in the cases
of the BCRB and the EZZB, admit a closed form for rectangular maps; 
\item \noindent they allow to a) identify map features (e.g., shape, area,
etc.) influencing localization accuracy and b) quantify such an influence.
\end{enumerate}
\noindent Then, such bounds are evaluated and compared for simple
representative environments. Our results allow to assess the importance
of map-awareness in localization systems in the presence of noisy
observations. In particular, they evidence that: 
\begin{enumerate}
\item maps should be expected to play a significant role at low signal-to-noise
ratios (SNRs) and in the presence of obstructions; 
\item \noindent in some cases simplified map models can be adopted for localization
purposes, since map details have a negligible impact on estimation
accuracy. 
\end{enumerate}
The manuscript is organized as follows. In Section \ref{sec:scenario}
our reference scenario is illustrated and the map model is defined.
Performance bounds for map-aware localization are derived and evaluated
in specific scenarios in Sections \ref{sec:perf_bounds} and \ref{sec:num_res},
respectively. Finally, some conclusions are offered in Section \ref{sec:conc}.

\emph{Notations}: Matrices are denoted by upper-case bold letters,
vectors by lower-case bold letters and scalar quantities by italic
letters. The notation $\mathbb{E}_{\mathbf{r}}\left\{ \cdot\right\} $
denotes expectation with respect to the random vector $\mathbf{r}$;
$\trace\{\cdot\}$ denotes the trace of a square matrix; $\diag\{\cdot\}$
denotes a square matrix with the arguments on its main diagonal and
zeros elsewhere; $[\cdot]_{i,l}$ is the element on the $i$-th row
and $l$-th column of its argument. $\mathbf{I}_{N}$ denotes the
$N\times N$ identity matrix; $\mathbf{A}\succeq\mathbf{B}$ means
that the matrix $\mathbf{A}-\mathbf{B}$ is positive semi-definite.
The probability density function (pdf) of the random vector $\mathbf{R}$
evaluated at the point $\mathbf{r}$ is denoted as $f(\mathbf{r})$,
whereas $\mathcal{N}\left(\mathbf{r};\mathbf{m},\boldsymbol{\Sigma}\right)$
denotes the pdf of a Gaussian random vector having mean $\mathbf{m}$
and covariance matrix $\boldsymbol{\Sigma}$, evaluated at the point
$\mathbf{r}$. $\mathbb{I}_{\mathcal{S}}(\mathbf{x})$ denotes the
so called indicator function for the set $\mathcal{S}$ (it is equal
to 1 when $\mathbf{x}\in\mathcal{S}$ and zero otherwise).

\section{Reference Scenario and Map Modelling\label{sec:scenario}}

In the following we focus on the problem of localizing a single device,
called \emph{agent}, in a 2-D environment, i.e., of estimating its
position $\mathbf{p}\triangleq[x,y]^{T}$, in the presence of the
following information: a) a map representing the a priori knowledge
about the environment in which the localization system operates; b)
an observation vector $\mathbf{z}$ related to the true position $\mathbf{p}$
of the agent.

\subsection{Map Modelling\label{sub:scenario_map}}

In a Bayesian framework the map of a given environment can be modelled
for localization purposes as a pdf $f(\mathbf{p})$ of the random
variable $\mathbf{p}$, representing the agent position. Such a pdf
is characterized by a 2-D support $\mathcal{R}$ having a finite size
and consisting of the set of points in the environment not occupied
by obstructions (e.g., walls, buildings, etc). In the absence of other
prior knowledge, a natural choice for $f(\mathbf{p})$ is a simple
uniform model, i.e., 
\begin{equation}
f(\mathbf{p})=f(x,y)=\begin{cases}
1/\mathcal{A_{R}} & \mathbf{p}\in\mathcal{R}\\
0 & \text{elsewhere}
\end{cases}\label{eq:uniform_map}
\end{equation}
where $\mathcal{A_{R}}$ denotes the area of $\mathcal{R}$. This
model is referred to as \emph{uniform map} in the following and is
preferable to other, more detailed, prior models since it can be adopted
when only basic knowledge of the environment is available%
\footnote{More detailed prior modelling requires additional information both
for outdoor and indoor environments (e.g., the end use of rooms, the
authorization levels to access them, furniture disposition, etc).
Such information are often time-variant (e.g., the furniture may be
moved), agent-specific (e.g., authorization levels and human habits
may vary) and difficult to acquire. On the contrary, uniform maps
only require the knowledge of building maps.%
} (e.g., the floor plan of a building in indoor environments or satellite
photos in outdoor environments). An alternative statistical model
for maps, better suited to mathematical analysis in the BCRB case,
is a ``smoothed'' version of (\ref{eq:uniform_map}); this is obtained
modifying the uniform pdf in a narrow area around the edges of $\mathcal{R}$
in order to introduce a smooth transition to zero along the boundaries
of its support. Note that the adoption of a smoothed uniform model
leads to analytical results similar to those found with the uniform
model if the transition region is small with respect to the size of
$\mathcal{R}$.

As it will become clearer in the following, for a generic map, a detailed
description of the structure of its support $\mathcal{R}$ should
be provided to ease the formulation of the accuracy bounds referring
to the map itself. 

To begin, let us define, for each $(x,y)\in\mathbb{R}^{2}$, the sets
$\mathcal{R}_{h}(y)\triangleq\left\{ t|(t,y)\in\mathcal{R}\right\} $
and $\mathcal{R}_{v}(x)\triangleq\left\{ t|(x,t)\in\mathcal{R}\right\} $
representing the intersection between $\mathcal{R}$ and an horizontal
(\emph{h}) and vertical (\emph{v}) line identified by the abscissa
$x$ and ordinate $y$, respectively. Then, let: 
\begin{itemize}
\item $N_{h}(y)$ and $N_{v}(x)$ denote the number of connected components
of $\mathcal{R}_{h}(y)$ and $\mathcal{R}_{v}(x)$, respectively; 
\item $\mathcal{N}_{h}^{o}(y)$ and $\mathcal{N}_{v}^{o}(x)$ denote the
set of the \emph{odd} values $\left\{ 1,3,...,2N_{h}(y)-1\right\} $
and $\left\{ 1,3,...,\right.$ $\left.2N_{v}(x)-1\right\} $, respectively; 
\item $w_{n}(y)$ and $h_{m}(x)$, with $n\in\mathcal{N}_{h}^{o}(y)$ and
$m\in\mathcal{N}_{v}^{o}(x)$, denote the length of the connected
components of $\mathcal{R}_{h}(y)$ and $\mathcal{R}_{v}(x)$ respectively; 
\item $\mathcal{N}_{h}^{e}(y)$ and $\mathcal{N}_{v}^{e}(x)$ denote the
set of the \emph{even} values $\left\{ 2,4,...,2\left(N_{h}(y)-1\right)\right\} $
and $\left\{ 2,4,...,\right.$ $\left.2\left(N_{v}(x)-1\right)\right\} $,
respectively. 
\item $\triangle w_{n}(y)$ and $\triangle h_{m}(x)$, with $n\in\mathcal{N}_{h}^{e}(y)$
and $m\in\mathcal{N}_{v}^{e}(x)$, denote the length of the connected
components of $\mathcal{R}_{h}^{c}(y)$ and $\mathcal{R}_{v}^{c}(x)$
respectively (note that $\mathcal{R}_{h}^{c}(y)$ and $\mathcal{R}_{v}^{c}(x)$
represent the complementary sets of $\mathcal{R}_{h}(y)$ and $\mathcal{R}_{v}(x)$
with respect to $\mathbb{R}$); in other words, they denote the horizontal
and vertical size of map obstructions, respectively;
\item $\mathcal{X}$ and $\mathcal{Y}$ denote the projection of $\mathcal{R}$
on the $x$ and $y$ axis, respectively; 
\end{itemize}
\begin{figure*}
\centering{}\includegraphics[width=12cm]{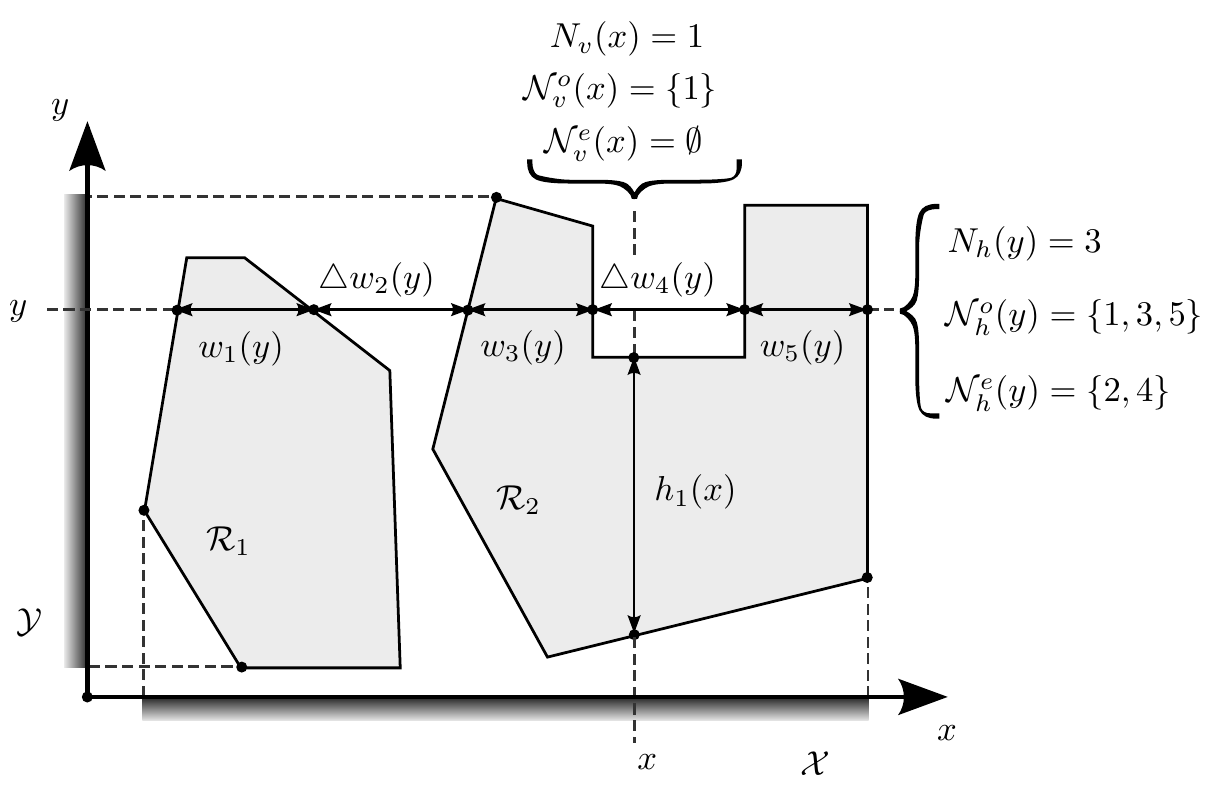}\caption{Support region $\mathcal{R}=\mathcal{R}_{1}\bigcup\mathcal{R}_{2}$
of a non-convex uniform map (the notation introduced in Sec. \ref{sub:scenario_map}
has been adopted). Note that $\mathcal{R}$ (grey region) represents
the area where the agent has non-zero probability to lie, so that
everything else has to be considered as an ``obstruction'' (see
(\ref{eq:uniform_map})). \label{fig:non_convex_generic_shape}}
\end{figure*}

The graphical meaning of these parameters for a specific uniform map
is exemplified by Fig. \ref{fig:non_convex_generic_shape}.

\subsection{Observation Modelling}

In the following, we assume that the localization system is able to
acquire a set of noisy observations, collected in a vector $\mathbf{z}$
and related to the agent position $\mathbf{p}$ according to a specific
statistical model. Note that, on the one hand, the results illustrated
in Sec. \ref{sub:bcrb} hold for a general statistical model relating
$\mathbf{z}$ to $\mathbf{p}$; on the other hand, in Sec. \ref{sub:zzb}
and \ref{sub:wwb}, it is assumed that 
\begin{equation}
f(\mathbf{z}|\mathbf{p})=\mathcal{N}\left(\mathbf{z};\mathbf{p},\boldsymbol{\Sigma}\right)\label{eq:obs_model_gaussian}
\end{equation}
for mathematical convenience, where $\mathbf{z}\in\mathbb{R}^{2}$
and $\boldsymbol{\Sigma}\triangleq\diag\left\{ \sigma_{x}^{2},\sigma_{y}^{2}\right\} $
(i.e., $\mathbf{z}$ is affected by uncorrelated noise). It is important
to point out that:
\begin{enumerate}
\item The model (\ref{eq:obs_model_gaussian}) is purposely abstract and
does not explicitly refer to a specific localization technique (e.g.,
TOA or received signal strength (RSS)) or to particular propagation
conditions. Generally speaking, it is suitable to describe the position
estimate $\mathbf{z}$ generated by a map-unaware and unbiased localization
algorithm; in addition, it leads to useful bounds which unveil the
impact of \emph{map} modelling (instead of that of \emph{observation}
modelling) on estimation performance; readers interested in an in-depth
analysis on the impact of observation modelling can refer to \cite{Qi2002,Larsson2004,Gustafsson2005,Qi2006,Gezici2005,Gezici2009,Shen2010,Shen2010a}.
\item In principle, the bounds derived in this manuscript can be extended
to the case of correlated noise, accounting for the presence of an
off-diagonal term $\sigma_{xy}\neq0$ in $\boldsymbol{\Sigma}$. However,
the presence of noise correlation is neglected in the following derivations
since: a) the magnitude of $\sigma_{xy}$ in a real world system depends
on several parameters (the type of measurements processed by the system,
the multi-lateration or angulation technique employed, the properties
of the propagation channel, etc) and cannot be easily assessed; b)
our research work evidenced that the presence of noise correlation
does not provide significant additional insights on the impact of
map-awareness on performance bounds.
\end{enumerate}

\section{Performance Bounds for Map-Aware Position Estimators\label{sec:perf_bounds}}

In this Section various bounds about the accuracy of map-aware position
estimation are derived and their implications are analysed.

\subsection{Bayesian Cramer-Rao Bounds\label{sub:bcrb}}

Given an estimator $\hat{\mathbf{p}}(\mathbf{z})$ of $\mathbf{p}$
based on the observation vector $\mathbf{z}$, the BCRB establishes
that the Bayesian mean square error (BMSE) matrix 
\[
\boldsymbol{\Lambda}\left(\hat{\mathbf{p}}(\mathbf{z})\right)\triangleq\mathbb{E}_{\mathbf{z},\mathbf{p}}\left\{ \left(\hat{\mathbf{p}}(\mathbf{z})-\mathbf{p}\right)\left(\hat{\mathbf{p}}(\mathbf{z})-\mathbf{p}\right)^{T}\right\} 
\]
satisfies $\boldsymbol{\Lambda}\left(\hat{\mathbf{p}}(\mathbf{z})\right)\succeq\mathbf{J}^{-1}$
\cite{treesIII,kay}, where $\mathbf{J}$ is the Bayesian Fisher information
matrix (BFIM) associated to the statistical models employed by $\hat{\mathbf{p}}(\mathbf{z})$.
This result entails that, for 2-D localization: 
\begin{equation}
\mathbf{E}\triangleq\diag\left\{ E_{x},E_{y}\right\} \succeq\mathbf{B}\triangleq\diag\left\{ B_{x},B_{y}\right\} \label{eq:bcrb}
\end{equation}
where $E_{x}\triangleq\left[\boldsymbol{\Lambda}\left(\hat{\mathbf{p}}(\mathbf{z})\right)\right]_{1,1}$,
$E_{y}\triangleq\left[\boldsymbol{\Lambda}\left(\hat{\mathbf{p}}(\mathbf{z})\right)\right]_{2,2}$,
$B_{x}\triangleq\left[\mathbf{J}^{-1}\right]_{1,1}$ and $B_{y}\triangleq\left[\mathbf{J}^{-1}\right]_{2,2}$.
It is not difficult to show that the BFIM $\mathbf{J}$ can be put
in the form (e.g., see \cite[p. 183, eq. (75)]{treesIII}) 
\begin{equation}
\mathbf{J}=\mathbf{J}_{\mathbf{z|p}}+\mathbf{J}_{\mathbf{p}}\label{eq:j_somma}
\end{equation}
where 
\begin{equation}
\mathbf{J}_{\mathbf{z|p}}\triangleq\mathbb{E}_{\mathbf{z},\mathbf{p}}\left\{ -\frac{\partial}{\partial\mathbf{p}}\left[\frac{\partial}{\partial\mathbf{p}}\ln f(\mathbf{z}|\mathbf{p})\right]^{T}\right\} \label{eq:bfim_cond}
\end{equation}
and 
\begin{equation}
\mathbf{J}_{\mathbf{p}}\triangleq\mathbb{E}_{\mathbf{p}}\left\{ -\frac{\partial}{\partial\mathbf{p}}\left[\frac{\partial}{\partial\mathbf{p}}\ln f(\mathbf{p})\right]^{T}\right\} \label{eq:bfim_apriori}
\end{equation}
represent the contributions originating from the observation vector
(i.e., the so called \emph{conditional} information) and that due
to \emph{a priori} information, respectively. The last decomposition
allows us to analyse the two above mentioned sources of information
in a separate fashion. Therefore, in the following, we mainly focus
on the evaluation of the matrix $\mathbf{J}_{\mathbf{p}}$ (\ref{eq:bfim_apriori}),
since this term unveils the contribution of map information in the
estimation of agent position. As already mentioned in Sec. \ref{sub:scenario_map}
a smoothed uniform map model is adopted in the derivation of $\mathbf{J}_{\mathbf{p}}$
(analytical details are provided in Appendix \ref{apd:proofs_bfim})
so that its pdf $f(\mathbf{p})$, unlike its (discontinuous) uniform
counterpart (\ref{eq:uniform_map}), satisfies the regularity condition\cite{kay}
\begin{equation}
\mathbb{E}_{\mathbf{p}}\left\{ \frac{\partial}{\partial\mathbf{p}}\ln f(\mathbf{p})\right\} =\mathbf{0}\label{eq:regularity_condition}
\end{equation}
required for the evaluation of the Bayesian Fisher information. Moreover
it is assumed that:
\begin{enumerate}
\item the smoothing of $f(\mathbf{p})$ is described by a smoothing function
$s(t)$, which is required to be a continuous and differentiable pdf
for which the a priori Fisher information (FI) $\mathrm{J}_{s}\triangleq\mathbb{E}_{t}\left\{ \left(\frac{\partial\ln s\left(t\right)}{\partial t}\right)^{2}\right\} $
can be evaluated; 
\item the smoothing of $f(\mathbf{p})$ along the $x$ axis is independent
from that along the $y$ axis. 
\end{enumerate}
In Appendix \ref{apd:proofs_bfim} it is proved that, given the assumptions
illustrated above, $\mathbf{J}_{\mathbf{p}}$ (\ref{eq:bfim_apriori})
can be approximated as 
\begin{equation}
\mathbf{J_{p}}\simeq\frac{\mathrm{J}_{s}}{\mathcal{A_{R}}}\diag\left\{ \int_{\mathcal{Y}}\sum_{n\in\mathcal{N}_{h}^{o}(y)}\frac{dy}{w_{n}(y)},\int_{\mathcal{X}}\sum_{m\in\mathcal{N}_{v}^{o}(x)}\frac{dx}{h_{m}(x)}\right\} \label{eq:bfim_map_generic_mult_intervals}
\end{equation}
Note that such an expression is independent from the observation model
and depends from the smoothing function through its FI $\mathrm{J}_{s}$
only%
\footnote{In Section \ref{sub:zzb} the value to be assigned to $\mathrm{J}_{s}$
is derived for the case of rectangular maps.%
}. Given $\mathbf{J_{p}}$, the computation of the BCRB requires the
knowledge of $\mathbf{J}_{\mathbf{z|p}}$ (\ref{eq:bfim_cond}) (see
(\ref{eq:j_somma})); for its evaluation, instead of considering a
specific observation model, we just assume that the cross-information
about $x$ and $y$ is negligible, so that such a matrix can be put
in the form%
\footnote{The following expression becomes $\mathbf{J_{z|p}}=\diag\left\{ \sigma_{x}^{-1},\sigma_{y}^{-1}\right\} $
if the specific observation model (\ref{eq:obs_model_gaussian}) is
adopted; this result will be exploited later for comparison with other
bounds derived for the model (\ref{eq:obs_model_gaussian}).%
} 
\begin{equation}
\mathbf{J_{z|p}}=\diag\left\{ \mathrm{J}_{\mathbf{z}|x},\mathrm{J}_{\mathbf{z}|y}\right\} \label{eq:Jzp}
\end{equation}
Then, the BCRB associated with the pdf $f(\mathbf{p})$ can be put
in the form (see (\ref{eq:bfim_map_generic_mult_intervals}), (\ref{eq:Jzp})
and (\ref{eq:j_somma})) 
\begin{align}
\mathbf{E}\succeq\mathbf{B}\simeq\diag & \left\{ \left(\mathrm{J}_{\mathbf{z}|x}+\frac{\mathrm{J}_{s}}{\mathcal{A_{R}}}\int_{\mathcal{Y}}\sum_{n\in\mathcal{N}_{h}^{o}(y)}\frac{dy}{w_{n}(y)}\right)^{-1},\right.\notag\nonumber \\
 & \quad\left.\left(\mathrm{J}_{\mathbf{z}|y}+\frac{\mathrm{J}_{s}}{\mathcal{A_{R}}}\int_{\mathcal{X}}\sum_{m\in\mathcal{N}_{v}^{o}(x)}\frac{dx}{h_{m}(x)}\right)^{-1}\right\} \label{eq:bcrb_generic}
\end{align}
It can be shown that the last two expressions are exact, and can be
put in a closed form (not involving any integral), if $\mathcal{R}$
is the union of a set of disjoint rectangles (all having parallel
sides).

From (\ref{eq:bfim_map_generic_mult_intervals}) and (\ref{eq:bcrb_generic})
it can be inferred that: 
\begin{enumerate}
\item Map-aware estimators should be expected to be always more accurate
than their map-unaware counterparts; in fact $\mathbf{J_{p}}$ (\ref{eq:bfim_map_generic_mult_intervals})
is positive definite, so that the trace of the matrix $\mathbf{B}$
(\ref{eq:bcrb_generic}) (which bounds the accuracy of map-aware estimators)
is smaller than the trace of 
\[
\left(\mathbf{J_{z|p}}\right)^{-1}=\diag\left\{ \mathrm{J}_{\mathbf{z}|x}^{-1},\mathrm{J}_{\mathbf{z}|y}^{-1}\right\} 
\]
(see (\ref{eq:Jzp})) which represents the Cramer-Rao bound (CRB)
in the considered scenario (and bounds the accuracy of map-unaware
estimators). Note that this is a well-known result (the contribution
of a priori information is positive and lowers the BCRB), but may
be concealed by the complex analytical form of (\ref{eq:bcrb_generic}).
\item The BCRB (\ref{eq:bcrb_generic}) exhibits a complicated dependence
on the map properties and, in particular, a complex non linear dependence
on the map area $\mathcal{A_{R}}$; in fact, this parameter can be
related to the functions $\left\{ h_{m}(x)\right\} $ and $\left\{ w_{n}(y)\right\} $
appearing in the right-hand side (RHS) of (\ref{eq:bcrb_generic})
as 
\[
\mathcal{A_{R}}=\int_{\mathcal{Y}}\sum_{n\in\mathcal{N}_{h}^{o}(y)}w_{n}(y)dy=\int_{\mathcal{X}}\sum_{m\in\mathcal{N}_{v}^{o}(x)}h_{m}(x)dx\ .
\]
However, the following asymptotic result suggests that the BCRB is
expected to increase (and thus the accuracy of map-aware estimation
to worsen) as $\mathcal{A_{R}}$ gets larger: if $\mathcal{A_{R}}\rightarrow\infty$
the BCRB tends to the CRB (since the map support $\mathcal{R}\rightarrow\mathbb{R}^{2}$
and the contribution coming from prior information vanishes%
\footnote{If $\mathcal{R}\rightarrow\mathbb{R}^{2}$ the prior expressed by
(\ref{eq:uniform_map}) becomes ``improper'' (see \cite[Sec. 4.2]{Kass1996}).%
}) which always takes on a bigger value (see point 1).
\item Only the smallest elements in the set $\left\{ w_{n}(y)\right\} $
($\left\{ h_{m}(x)\right\} $) significantly contribute to the sum
$\vphantom{A_{A_{A_{A_{A_{A}}}}}}\sum_{n\in\mathcal{N}_{h}^{o}(y)}\frac{1}{w_{n}(y)}$
($\sum_{m\in\mathcal{N}_{v}^{o}(x)}\frac{1}{h_{m}(x)}$) appearing
in (\ref{eq:bfim_map_generic_mult_intervals}) and (\ref{eq:bcrb_generic})
and, consequently, may appreciably influence estimation accuracy;
in the limit, if $w_{n}(y)\rightarrow0$ ($h_{m}(x)\rightarrow0$)
for some $n$ and $y$ ($m$ and $x$), then the information about
$x$ ($y$) coordinate becomes infinite because the agent is constrained
to lie, for those values of $n$ and $y$ ($m$ and $x$), in a single
point of the map support. 
\item The matrices $\mathbf{J_{p}}$ (\ref{eq:bfim_map_generic_mult_intervals})
and $\mathbf{B}$ (\ref{eq:bcrb_generic}) can be easily put in a
closed form when $\left\{ h_{m}(x)\right\} $ and $\left\{ w_{n}(y)\right\} $
are step functions (i.e., the map is made by union of rectangles).
\end{enumerate}
Further insights can be obtained taking a specific class of maps into
consideration. In particular, here we focus on the case of a map without
obstructions, i.e., characterized by a support $\mathcal{R}$ such
that $\mathcal{N}_{h}^{e}(y)=\emptyset$ and $\mathcal{N}_{v}^{e}(x)=\emptyset$
$\forall x,y$ (note that convex maps belong to this class); under
these assumptions, it is easy to show that $\mathbf{J}_{\mathbf{p}}$
(\ref{eq:bfim_map_generic_mult_intervals}) can be simplified as 
\begin{equation}
\mathbf{J_{p}}\simeq\frac{\mathrm{J}_{s}}{\mathcal{A_{R}}}\diag\left\{ \int_{\mathcal{Y}}\frac{dy}{w_{1}(y)},\int_{\mathcal{X}}\frac{dx}{h_{1}(x)}\right\} \label{eq:bfim_maps_nogap}
\end{equation}
From the RHS of the last expression it can be easily inferred that:
\begin{figure*}[!t]

\normalsize
\setcounter{MYtempeqncnt}{\value{equation}}
\setcounter{equation}{13}

\begin{align}
\mathbf{E}\succeq\mathbf{Z}=\diag & \left\{ \frac{\sigma_{x}^{3}}{2\mathcal{\mathcal{A_{R}}}}\int_{\mathcal{Y}}\left[\sum_{n\in\mathcal{N}_{h}^{o}(y)}\zeta\left(\frac{w_{n}(y)}{\sigma_{x}}\right)+\sum_{n\in\mathcal{N}_{h}^{e}(y)}\zeta_{ov}\left(\frac{\triangle w_{n}(y)}{\sigma_{x}},\frac{w_{n-1}(y)}{\sigma_{x}},\frac{w_{n+1}(y)}{\sigma_{x}}\right)\right]dy,\right.\nonumber \\
 & \left.\frac{\sigma_{y}^{3}}{2\mathcal{A_{R}}}\int_{\mathcal{X}}\left[\sum_{m\in\mathcal{N}_{v}^{o}(x)}\zeta\left(\frac{h_{m}(x)}{\sigma_{y}}\right)+\sum_{m\in\mathcal{N}_{v}^{e}(x)}\zeta_{ov}\left(\frac{\triangle h_{m}(x)}{\sigma_{y}},\frac{h_{m-1}(x)}{\sigma_{y}},\frac{h_{m+1}(x)}{\sigma_{y}}\right)\right]dx\right\} \label{eq:zzb_generic}
\end{align}
\setcounter{equation}{\value{MYtempeqncnt}}
\hrulefill
\vspace*{4pt}
\end{figure*}

\begin{enumerate}
\item Even in this case the dependence of $\mathbf{J_{p}}$ from $\mathcal{A_{R}}$
is complicated. However, analysing (\ref{eq:bfim_maps_nogap}), it
is easy to note that if $\mathcal{R}\rightarrow\mathbb{R}^{2}$ (so
that $\mathcal{A_{R}}\rightarrow\infty$) the elements of $\mathbf{J_{p}}$
tend to zero, so that the impact of prior information vanishes, as
already mentioned above.
\item The amount of information concerning the agent position along the
$x$ ($y$) axis direction is proportional to $\int_{\mathcal{Y}}\tilde{w}(y)dy$
($\int_{\mathcal{X}}\tilde{h}(x)dx$), where $\tilde{w}(y)\triangleq1/w_{1}(y)$
($\tilde{h}(x)\triangleq1/h_{1}(x)$); the last integral can be interpreted
as the area of a ``virtual map'' characterized by the same projection
$\mathcal{Y}$ ($\mathcal{X}$) as the real map, but whose width (height)
at each point $y\mathcal{\in Y}$ ($x\mathcal{\in X}$) is the reciprocal
of the width (height) of the real map.
\item When prior information dominates over conditional information, the
map support $\mathcal{R}$ minimizing the trace of the BCRB under
the constraint of a constant area $\mathcal{A_{R}}$ has a square
shape. In fact, if $\mathcal{R}$ is a rectangle whose sides have
lengths $L_{x}$ and $L_{y}$, it is easy to show that $\mathbf{J_{p}}=\frac{\mathrm{J}_{s}}{\mathcal{A_{R}}}\diag\left\{ \frac{L_{y}}{L_{x}},\frac{L_{x}}{L_{y}}\right\} =\mathrm{J}_{s}\diag\left\{ \frac{1}{L_{x}^{2}},\frac{1}{L_{y}^{2}}\right\} $;
then, if we set $L_{x}=\gamma L$, $L_{y}=\frac{L}{\gamma}$, where
$\gamma$ and $L$ denote two real positive parameters, the trace
of the inverse of (\ref{eq:bfim_maps_nogap}) can be put in the form
$\frac{L^{2}}{\mathrm{J}_{s}}\left(\gamma^{2}+1/\gamma^{2}\right)$
and is minimized for $\gamma=1$ (i.e., for $L_{x}=L_{y}$). 
\item In (\ref{eq:bfim_maps_nogap}) the only contribution coming from the
smoothing function $s(t)$ is represented by the factor $\mathrm{J}_{s}$. 
\end{enumerate}
Further comments about the BCRB (\ref{eq:bcrb_generic}) will be provided
in Section \ref{sec:num_res}, where this bound is computed for some
specific maps.

Finally, it is important to point out that: a) in localization problems
BCRB's often provide useful insights \cite{Shen2010a,Shen2010b,Shen2010},
but these bounds are usually loose for low SNR conditions \cite{Seidman1970,Chazan1975},
i.e., when a priori information (the map in this context) plays a
critical role due to the poor quality of observations; b) the BCRB
analysis requires the adoption of the smoothed uniform pdf model for
prior information (see Section \ref{sub:scenario_map}). 

These considerations motivate the search for other bounds and, in
particular, for the EZZB, which is usually tighter than the BCRB at
low SNRs \cite{Ziv1969,Chazan1975,Bell1997,Nicholson1998,Montorsi_BCRB_CHANNEL}
and does not require (\ref{eq:regularity_condition}) to hold (so
that the uniform model (\ref{eq:uniform_map}) can be employed as
it is); this is further discussed in the following section.

\subsection{Zik-Zakai Bounds\label{sub:zzb}}

Similarly to (\ref{eq:bcrb}), the EZZB for 2D localization can be
expressed as 
\begin{equation}
\mathbf{\mathbf{E}}\triangleq\diag\left\{ E_{x},E_{y}\right\} \succeq\mathbf{Z}\triangleq\diag\left\{ Z_{x},Z_{y}\right\} \label{eq:zzb}
\end{equation}
where (see Appendix \ref{apd:proofs_zzb}) 
\begin{equation}
Z_{\nu}\triangleq\frac{1}{2}\iint_{\mathcal{P}_{\nu}}\left[f\left(\boldsymbol{\rho}\right)+f\left(\boldsymbol{\rho}+h\mathbf{e}_{\nu}\right)\right]P_{\text{\tiny{min}}}^{\mathbf{z}}\left(\boldsymbol{\rho},\boldsymbol{\rho}+h\mathbf{e}_{\nu}\right)h\, d\boldsymbol{\rho}dh\label{eq:zzb_def}
\end{equation}
$\nu\in\{x,y\}$, $\mathbf{e}_{x}\triangleq[1,0]^{T}$, $\mathbf{e}_{y}\triangleq[0,1]^{T}$,
$P_{\text{\tiny{min}}}^{\mathbf{z}}(\boldsymbol{\rho},\boldsymbol{\rho}+h\mathbf{e}_{\nu})$
represents the error probability referring to the likelihood ratio
test in a binary detection problem involving the hypotheses 
\[
H_{0}:\:\mathbf{p}=\boldsymbol{\rho}
\]
\[
H_{1}:\:\mathbf{p}=\boldsymbol{\rho}+h\mathbf{e}_{\nu}
\]
 and the observation $\mathbf{z}$; finally, the integration domain
is 
\[
\mathcal{P}_{\nu}\triangleq\left\{ \left(h,\boldsymbol{\rho}\right):h\geq0\wedge f\left(\boldsymbol{\rho}\right)>0\wedge f\left(\boldsymbol{\rho}+h\mathbf{e}_{\nu}\right)>0\right\} \subset\mathbb{R}^{3}
\]
The EZZB (\ref{eq:zzb})-(\ref{eq:zzb_def}) deserves the following
comments: 
\begin{enumerate}
\item its formulation is obtained from the standard formulation (see \cite[Eq. (32)]{Bell1997})
removing the so-called valley-filling function and the maximization
operator, as shown in Appendix \ref{apd:proofs_zzb}; these modifications
simplify the derivation of the bound at the price of a small reduction
in its tightness;
\item it exhibits a complicated dependence on the observation model, which
comes into play in the evaluation of the probability $P_{\text{\tiny{min}}}^{\mathbf{z}}\left(\boldsymbol{\rho},\boldsymbol{\rho}+h\mathbf{e}_{\nu}\right)$;
\item it cannot be put in a form similar to (\ref{eq:j_somma}), so that,
generally speaking, the contribution coming from conditional information
(i.e., observations) cannot be easily separated from that associated
with a priori information (map awareness in this case). 
\end{enumerate}
As shown in Appendix \ref{apd:proofs_zzb}, for a uniform map and
the observation model (\ref{eq:obs_model_gaussian}) the EZZB (\ref{eq:zzb_def})
can be put in the form (\ref{eq:zzb_generic}) shown at the top of
this page where 
\addtocounter{equation}{1} 
\begin{equation}
\zeta(\rho)\triangleq\int_{0}^{\rho}(\rho-u)u\erfc\left(\frac{u}{2\sqrt{2}}\right)du\label{eq:f_rho_def}
\end{equation}
\begin{multline}
\zeta_{ov}\left(\rho_{\triangle},\rho_{1},\rho_{2}\right)\triangleq\\
\int_{\rho_{\triangle}}^{\rho_{\triangle}+\rho_{2}}(u-\rho_{\triangle})u\erfc\left(\frac{u}{2\sqrt{2}}\right)du+\\
\int_{\rho_{\triangle}+\rho_{2}}^{\rho_{\triangle}+\rho_{1}}\rho_{2}u\erfc\left(\frac{u}{2\sqrt{2}}\right)du+\\
\int_{\rho_{\triangle}+\rho_{1}}^{\rho_{\triangle}+\rho_{1}+\rho_{2}}(\rho_{\triangle}+\rho_{1}+\rho_{2}-u)u\erfc\left(\frac{u}{2\sqrt{2}}\right)du\label{eq:f_gap_def}
\end{multline}
An accurate analysis of the EZZB (\ref{eq:zzb_generic}) and a comparison
with the BCRB (\ref{eq:bcrb_generic}) lead to the following conclusions: 
\begin{enumerate}
\item The EZZB (\ref{eq:zzb_generic}) exhibits a similar dependence as
the BCRB (\ref{eq:bcrb_generic}) in terms of the quantities $\{w_{n}(y)\}$,
$\{h_{m}(y)\}$. In fact, if $\{w_{n}(y)\}$ and/or $\{h_{m}(x)\}$
increases, the trace of matrix $\mathbf{Z}$ in (\ref{eq:zzb_generic})
gets larger, since $\zeta$ (\ref{eq:f_rho_def}) is a monotonically
increasing function.
\item In the same environment the EZZB (\ref{eq:zzb_generic}) takes on
larger values than the BCRB (\ref{eq:bcrb_generic}) and results to
be a bound tighter to the performance of optimal estimators. This
is due to the fact that the terms $\zeta$ appearing in the RHS of
(\ref{eq:zzb_generic}) give the EZZB approximately the same behaviour
of the BCRB; however, the EZZB also contains terms based on $\zeta_{ov}$
(\ref{eq:f_gap_def}), which is a positive valued function.
\item The EZZB (\ref{eq:zzb_generic}), unlike the BCRB, is influenced by
the shape of map obstructions (through the $\zeta_{ov}$ function
and the $\left\{ \triangle w_{n}(y)\right\} $, $\left\{ \triangle h_{m}(y)\right\} $
parameters).
\item The two terms of (\ref{eq:zzb_generic}) depending on the function
$\zeta_{ov}$ (\ref{eq:f_gap_def}) are influenced by the spatial
extensions $\left\{ \triangle w_{n}(y)\right\} $ and $\left\{ \triangle h_{m}(y)\right\} $
of the obstructions separating \emph{consecutive} map segments (which
are characterized by the lengths $\{w_{n-1}(y),w_{n+1}(y)\}$ and
$\{h_{m-1}(y),h_{m+1}(y)\}$, respectively); in particular, the impact
of obstructions is significant when $\triangle w_{n}(y)$ ($\triangle h_{m}(y)$)
is greater than one of $\{w_{n-1}(y),w_{n+1}(y)\}$ ($\{h_{m-1}(y),h_{m+1}(y)\}$).
\item The integrals in the RHS of (\ref{eq:zzb_generic}) can be easily
put in a closed form when $\left\{ h_{m}(x)\right\} $ and $\left\{ w_{n}(y)\right\} $
are step functions (i.e., the map is made by union of rectangles).
\end{enumerate}
Similarly to the BCRB, further insights can be obtained considering
the class of maps without obstructions. In this case the EZZB (\ref{eq:zzb_generic})
simplifies as 
\begin{figure*}[!t]

\normalsize
\setcounter{MYtempeqncnt}{\value{equation}}
\setcounter{equation}{21}

\begin{align}
\mathbf{E} & \succeq\mathbf{W}=\diag\left\{ \sup_{h_{x}\in\mathbb{R}}\frac{\left[h_{x}\exp\left(-\frac{h_{x}^{2}}{8\sigma_{x}^{2}}\right)\lambda_{x}\left(h_{x},\mathcal{R}\right)\right]^{2}}{2\mathcal{\mathcal{A_{R}}}\left[\lambda_{x}\left(h_{x},\mathcal{R}\right)-\exp\left(-\frac{h_{x}^{2}}{2\sigma_{x}^{2}}\right)\gamma_{x}\left(h_{x},\mathcal{R}\right)\right]},\sup_{h_{y}\in\mathbb{R}}\frac{\left[h_{y}\exp\left(-\frac{h_{y}^{2}}{8\sigma_{y}^{2}}\right)\lambda_{y}\left(h_{y},\mathcal{R}\right)\right]^{2}}{2\mathcal{\mathcal{A_{R}}}\left[\lambda_{y}\left(h_{y},\mathcal{R}\right)-\exp\left(-\frac{h_{y}^{2}}{2\sigma_{y}^{2}}\right)\gamma_{y}\left(h_{y},\mathcal{R}\right)\right]}\right\} \label{eq:wwb_generic}
\end{align}
\setcounter{equation}{\value{MYtempeqncnt}}
\hrulefill
\vspace*{4pt}
\end{figure*}
 
\begin{alignat}{1}
\mathbf{E}\succeq\mathbf{Z}=\frac{1}{2\mathcal{\mathcal{A_{R}}}}\diag & \left\{ \sigma_{x}^{3}\int_{\mathcal{Y}}\zeta\left(\frac{w_{1}(y)}{\sigma_{x}}\right)dy,\right.\nonumber \\
 & \:\:\:\left.\sigma_{y}^{3}\int_{\mathcal{X}}\zeta\left(\frac{h_{1}(x)}{\sigma_{y}}\right)dx\right\} \label{eq:zzb_maps_nogap}
\end{alignat}
It is also worth mentioning that, if the map support $\mathcal{R}$
can be expressed as the union of a set of disjoint rectangles (all
having parallel sides), the EZZB (\ref{eq:zzb_maps_nogap}) can be
simplified further, obtaining a closed-form bound. 

The result (\ref{eq:zzb_maps_nogap}) deserves the following comments: 
\begin{enumerate}
\item The structure of its RHS is similar to that of the equivalent BFIM
(\ref{eq:bfim_maps_nogap}). Despite the differences between the integrand
functions, our numerical results show that in this case both the EZZB
and the BCRB are tight and that the EZZB does not provide additional
hints on localization accuracy with respect to the BCRB (see Section
\ref{sec:num_res}).
\item The EZZB predicts the correct variance of map-aware estimation when
the SNR is very low whereas a specific assumption about the value
of $\mathrm{J}_{s}$ is required to obtain the same result in the
BCRB case. For instance, let us consider a uniform map whose support
$\mathcal{R}$ is a rectangle whose sides have lengths $L_{x}$ and
$L_{y}$; with model (\ref{eq:obs_model_gaussian}) and for $\sigma_{x}^{2}\rightarrow\infty$
and $\sigma_{y}^{2}\rightarrow\infty$ (i.e., for a SNR approaching
zero), the BMSE of the minimum mean square error (MMSE) estimator
tends to $\diag\left\{ L_{x}^{2}/12,L_{y}^{2}/12\right\} $. The EZZB
provides this exact result since the RHS of (\ref{eq:zzb_maps_nogap})
can be put in the closed form 
\begin{equation}
\mathbf{Z}=\diag\left\{ \frac{\sigma_{x}^{3}}{2L_{x}}\zeta\left(\frac{L_{x}}{\sigma_{x}}\right),\frac{\sigma_{y}^{3}}{2L_{y}}\zeta\left(\frac{L_{y}}{\sigma_{y}}\right)\right\} \label{eq:zzb_rectangle_map}
\end{equation}
and $\zeta(\rho)/(2\rho)\rightarrow\rho^{2}/12$ for a SNR $\rho$
approaching zero%
\footnote{This proof requires developing a Taylor expansion of the function
$\zeta(\rho)/(2\rho)$ around $\rho=0$. %
}. This result can be compared with its BCRB counterpart; in this case,
from (\ref{eq:bfim_maps_nogap}) it can be inferred that $\mathbf{B}\mathbf{\rightarrow}\diag\left\{ L_{x}^{2}/J_{s},L_{y}^{2}/J_{s}\right\} $,
so that $J_{s}=12$ should be selected to match the correct variance.
\item The EZZB (\ref{eq:zzb_maps_nogap}) confirms that map-aware estimators
should be expected to be more accurate than their map-unaware counterparts,
at least for maps without obstructions: for instance, if a rectangular
map of sides $L_{x}$, $L_{y}$ is considered, then (\ref{eq:zzb_rectangle_map})
is easily proved to be always smaller than $\diag\left\{ \sigma_{x}^{2},\sigma_{y}^{2}\right\} $
(because $0\leq\zeta(\rho)/(2\rho)<1$) which is the trivial EZZB%
\footnote{This result is obtained from (\ref{eq:zzb_rectangle_map}) after proving
that $\zeta(\rho)/(2\rho)\rightarrow1$ for $\rho\rightarrow+\infty$.%
} for the observation model (\ref{eq:obs_model_gaussian}) when the
map is unavailable (i.e., when no prior information is available).
This derives from the fact that a priori information lowers the EZZB
(similarly to what occurs in the case of the CRB and the BCRB); this
is a well-known result, but may not be evident in (\ref{eq:zzb_maps_nogap}).
\item Similarly to the BCRB, the dependence on the map area is complicated;
however, it should be expected that the elements of $\mathbf{Z}$
get larger when $\mathcal{\mathcal{A_{R}}}$ increases as suggested
by the following asymptotic result: when $\mathcal{A_{R}}\rightarrow\infty$
the contribution coming from a priori information vanishes and the
EZZB (\ref{eq:zzb_rectangle_map}) tends to $\diag\left\{ \sigma_{x}^{2},\sigma_{y}^{2}\right\} $,
i.e., to the map-unaware EZZB, which is larger (see point 3). 
\end{enumerate}
As already mentioned above, the EZZB allows us to understand the role
played by some map features, which do not have any impact on the BCRB.
However, our numerical results evidence that, for complex maps and,
in particular, at very low SNRs, this bound turns out to be somewhat
far from the performance offered by a MMSE estimator. This has motivated
the derivation of the WWB \cite{Weiss1985}, which can be tighter
than the EZZB at very low-SNRs.

\subsection{Wess-Weinstein Bounds\label{sub:wwb}}

Similarly to the BCRB and the EZZB (see (\ref{eq:bcrb}) and (\ref{eq:zzb}),
respectively), the WWB, for 2-D localization, can be expressed as
\begin{equation}
\mathbf{E}\succeq\mathbf{W}\triangleq\diag\left\{ W_{x},W_{y}\right\} \label{eq:wwb}
\end{equation}
where \cite{Ben-Haim2008} 
\begin{equation}
W_{\nu}\triangleq\sup_{h_{\nu}\in\mathbb{R}}\frac{\left(h_{\nu}\mathbb{E}_{\mathbf{z},\mathbf{p}}\left\{ L^{\frac{1}{2}}\left(\mathbf{z};\mathbf{p}^{+},\mathbf{p}\right)\right\} \right)^{2}}{\mathbb{E}_{\mathbf{z},\mathbf{p}}\left\{ \left[L^{\frac{1}{2}}\left(\mathbf{z};\mathbf{p}^{+},\mathbf{p}\right)-L^{\frac{1}{2}}\left(\mathbf{z};\mathbf{p}^{-},\mathbf{p}\right)\right]^{2}\right\} }\label{eq:wwb_def}
\end{equation}
and
\[
\mathbf{p}^{+}\triangleq\mathbf{p}+h_{\nu}\mathbf{e}_{\nu}
\]
\[
\mathbf{p}^{-}\triangleq\mathbf{p}-h_{\nu}\mathbf{e}_{\nu}
\]
are functions of the $h_{\nu}$ parameter; moreover 
\begin{equation}
L\left(\mathbf{z};\mathbf{p}_{1},\mathbf{p}_{2}\right)\triangleq\frac{f(\mathbf{z},\mathbf{p}_{1})}{f(\mathbf{z},\mathbf{p}_{2})}=\frac{f(\mathbf{z}|\mathbf{p}_{1})f(\mathbf{p}_{1})}{f(\mathbf{z}|\mathbf{p}_{2})f(\mathbf{p}_{2})}\label{eq:likelihood_ratio}
\end{equation}
is a likelihood ratio and $\nu\in\{x,y\}$. Note that (\ref{eq:wwb_def})
can be obtained from \cite[Eqs. (7)-(8)]{Ben-Haim2008} setting $N=2$,
$s_{1}=s_{2}=\frac{1}{2}$ for the WWB optimization parameters%
\footnote{These choices make the bound tighter, as pointed out in \cite{Weiss1985}. %
} and $\mathbf{H}=\diag\left\{ h_{x},h_{y}\right\} $, since we are
interested in a bound on the variance (not on the co-variance) of
the estimates of $\mathbf{p}$.

As shown in Appendix \ref{apd:proofs_wwb}, for a uniform map and
the observation model (\ref{eq:obs_model_gaussian}) the WWB (\ref{eq:wwb_def})
can be put in the form (\ref{eq:wwb_generic}) shown at the top of
the next page,
\addtocounter{equation}{1} where $h_{x}$, $h_{y}$ denote the parameters to be optimized to
make the bound tighter, 
\begin{multline}
\lambda_{x}\left(h_{x},\mathcal{R}\right)\triangleq\int\mathbb{I}_{\mathcal{R}}(\mathbf{p})\mathbb{I}_{\mathcal{R}}(\mathbf{p}+h_{x}\mathbf{e}_{x})d\mathbf{p}\\
\shoveleft=\int_{\mathcal{Y}}\sum_{n\in\mathcal{N}_{h}^{o}(y)}\omega\left(w_{n}(y),h_{x}\right)dy+\\
\int_{\mathcal{Y}}\sum_{n\in\mathcal{N}_{h}^{e}(y)}\omega_{ov}\left(\triangle w_{n}(y),w_{n-1}(y),w_{n+1}(y),h_{x}\right)dy\label{eq:lambda_def}
\end{multline}
and 
\begin{align}
\gamma_{x}\left(h_{x},\mathcal{R}\right) & \triangleq\int\mathbb{I}_{\mathcal{R}}(\mathbf{p})\mathbb{I}_{\mathcal{R}}(\mathbf{p}+h_{x}\mathbf{e}_{x})\mathbb{I}_{\mathcal{R}}(\mathbf{p}-h_{x}\mathbf{e}_{x})d\mathbf{p}\nonumber \\
 & =2\int_{\mathcal{Y}}\sum_{n\in\mathcal{N}_{h}^{o}(y)}\omega\left(\frac{1}{2}w_{n}(y),h_{x}\right)dy\label{eq:gamma_def}
\end{align}
are functions of the parameters $h_{x}$, $h_{y}$ and of the geometrical
features of the map support $\mathcal{R}$ (dual expressions hold
for $\lambda_{y}\left(h_{y},\mathcal{R}\right)$ and $\gamma_{y}\left(h_{y},\mathcal{R}\right)$);
finally 
\begin{equation}
\omega\left(w,h_{x}\right)\triangleq(w-h_{x})\mathbb{I}_{[0;w]}(h_{x})\label{eq:omega_def}
\end{equation}
\begin{multline}
\omega_{ov}\left(\triangle w,w_{1},w_{2},h\right)\triangleq\\
\begin{cases}
h-\triangle w & h\in[\triangle w;\triangle w+w_{2}]\\
w_{2} & h\in[\triangle w+w_{2};\triangle w+w_{1}]\\
w_{tot}-h & h\in[\triangle w+w_{1};w_{tot}]
\end{cases}\label{eq:omega_ov_def}
\end{multline}
\[
w_{tot}\triangleq\triangle w+w_{1}+w_{2}\ .
\]
Note that the structure of \textbf{$\lambda_{x}\left(h_{x},\mathcal{R}\right)$}
(\ref{eq:lambda_def}) is similar to that of some terms contained
in the RHS of the EZZB (\ref{eq:zzb_generic}); however, in the WWB
the noise variances appear in some exponential functions (\ref{eq:wwb_generic})
only.

Unfortunately, it is hard to infer from (\ref{eq:wwb_generic}) any
conclusion about the dependence of the WWB on the features of a given
map; this is mainly due to the presence of the $\sup\left(\cdot\right)$
operators and to the discontinuous behaviour of the functions $\omega\left(\cdot\right)$
and $\omega_{ov}\left(\cdot\right)$. Despite this, the WWB may play
an important role, since, as shown in Section \ref{sec:num_res} for
specific maps, it is tighter than the BCRB and the EZZB at low-SNRs.

In the following Section the general bounds provided above are evaluated
for specific maps, so that some additional insights about the ultimate
accuracy achieved by map-aware estimators can be obtained.

\section{Numerical Results\label{sec:num_res}}

\begin{figure}
\centering{}\includegraphics[width=3in]{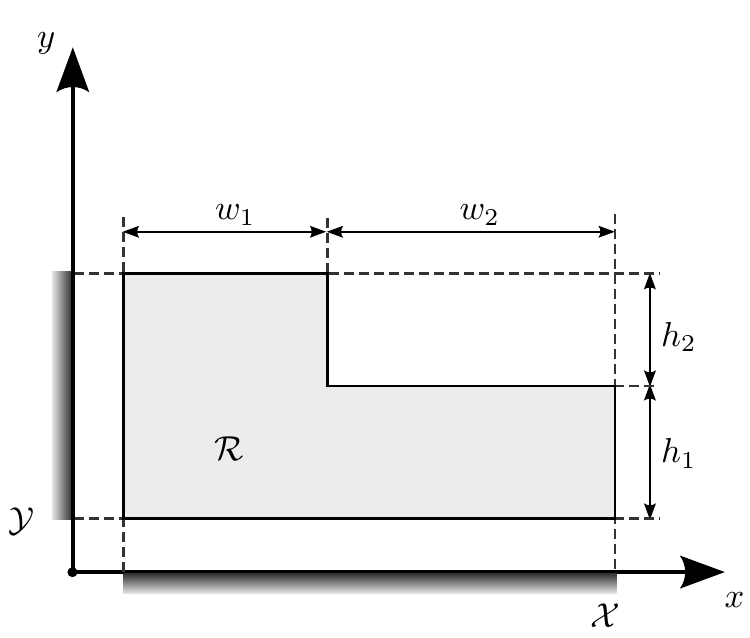}\caption{Support region $\mathcal{R}$ of a ``L-shaped'' map (map \#2).\label{fig:map_2d_lshape}}
\end{figure}

\begin{figure}
\centering{}\includegraphics[width=3in]{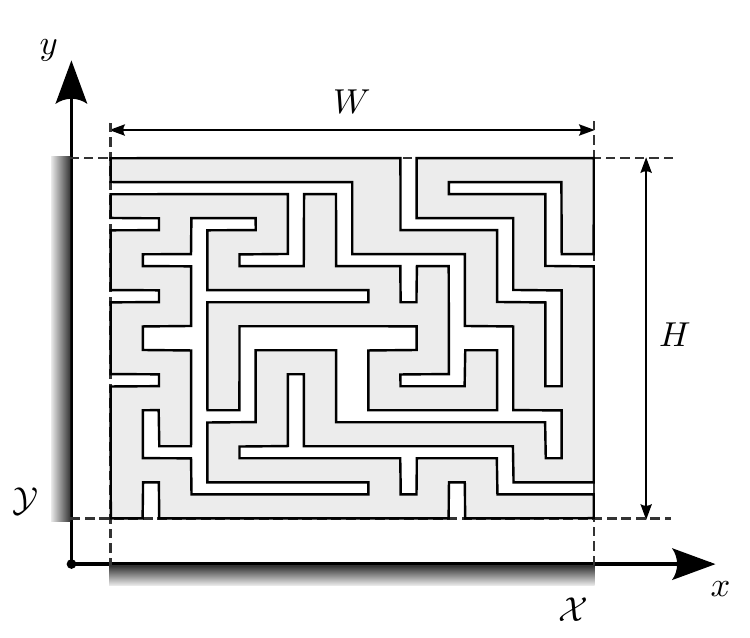}\caption{Support region $\mathcal{R}$ of a map modelling the floor of a big
building (map \#3).\label{fig:map_2d_indoor_building}}
\end{figure}

The bounds illustrated in the previous Section have been evaluated
for the following 3 specific maps: 
\begin{enumerate}
\item map \#1: a unidimensional map whose support $\mathcal{R}$ consists
of $N_{r}=2$ disjoint segments (denoted $\mathcal{R}_{1}$ and $\mathcal{R}_{2}$
in the following) having the same length $w$ and separated by $\triangle x$
meters; 
\item map \#2: the ``L-shaped'' 2-D map shown in Fig. \ref{fig:map_2d_lshape}
and fully described by the set $\left\{ w_{1},w_{2},\right.$ $\left.h_{1},h_{2}\right\} $
of geometrical parameters; 
\item map \#3: the 2-D map shown in Fig. \ref{fig:map_2d_indoor_building};
note that such a map models the floor of a large building and is quite
complex, so that its geometrical parameters are not given and analytical
bounds are not evaluated for it; instead, some numerical results generated
by means of computer simulations will be illustrated. 
\end{enumerate}
It is important to point out that maps \#2 and \#3 are examples of
2-D \emph{rectangular} uniform maps, i.e., each of them is characterised
by a support $\mathcal{R}$ that can be represented as the union of
non-overlapping rectangles, all having parallel sides; such non-overlapping
rectangles form the so called \emph{rectangular covering} of the map
support $\mathcal{R}$ \cite{Chaiken1981,Liou1990,Wu1994}. Note that
any map characterised by a support of \emph{arbitrary }shape can be
always approximated, up to an arbitrary precision, by a rectangular
map if a proper covering, containing a sufficiently large number of
rectangles, is selected. For this reason, maps \#2 and \#3 are relevant
and practical examples. Map \#1 can be also deemed relevant from a
practical point of view since it can be thought of as a transversal
slice of a 2-D map (consisting, for instance, of two road lanes or
two rooms facing each other); in addition, such a simple scenario
allows to acquire more insights than maps \#2 and \#3, as shown later.

In all cases the observation model (\ref{eq:obs_model_gaussian})
has been adopted (with $\boldsymbol{\Sigma}=\sigma^{2}$ and $\boldsymbol{\Sigma}=\sigma^{2}\mathbf{I}_{2}$
in the 1-D and 2-D scenarios, respectively) and the tightness of the
bounds has been assessed comparing them with the root mean square
error (RMSE) performance of optimal estimators evaluated by means
of extensive computer simulations%
\footnote{The software developed to generate our numerical results is available
at \texttt{\scriptsize{http://frm.users.sf.net/publications.html}}
in accordance with the philosophy of reproducible research standard
\cite{Vandewalle2009}.%
}. In addition, in accordance with the theoretical arguments of Paragraph
\ref{sub:zzb} (see the comments to (\ref{eq:zzb_maps_nogap})), $J_{s}=12$
has been selected in the evaluation of all the BCRB results illustrated
in this Section. Finally, in our simulations $N_{runs}=10^{4}$ realizations
of the random parameter $\mathbf{p}$ have been generated according
to a uniform distribution over the map support $\mathcal{R}$; then
Gaussian noise samples (characterized by $\sigma=3\meter$ if not
explicitly stated) have been added to generate the noisy observations,
according to (\ref{eq:obs_model_gaussian}).

\subsection{Map \#1\label{sub:num_res_map1}}

\begin{figure}
\centering{}\includegraphics[width=3.5in]{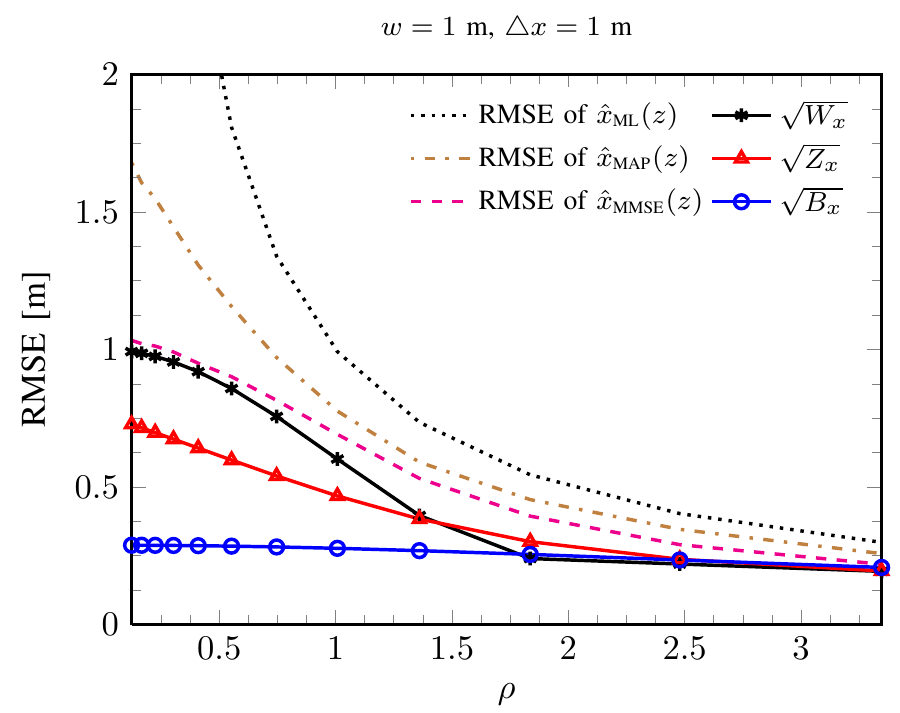}\caption{RMSE versus the SNR $\rho$ for the MMSE (\ref{eq:MMSE_estimator_map_1}),
MAP (\ref{eq:MAP_estimator_map_1}) and ML estimators. Map \#1, $w=1\meter$
and $\triangle x=1\meter$ are assumed. The BCRB (\ref{eq:map1_bcrb}),
EZZB (\ref{eq:map1_zzb}) and WWB (\ref{eq:map1_wwb}) are shown for
comparison.\label{fig:map1_vs_noise}}
\end{figure}

\begin{figure}
\centering{}\includegraphics[width=3.5in]{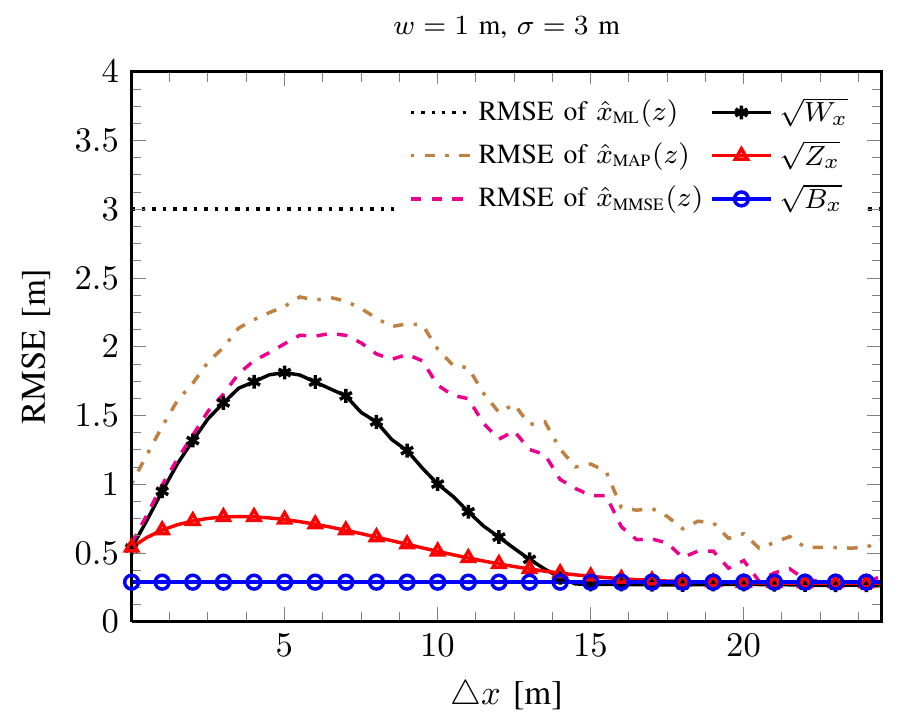}\caption{RMSE versus $\triangle x$ for the MMSE (\ref{eq:MMSE_estimator_map_1}),
MAP (\ref{eq:MAP_estimator_map_1}) and ML estimators. Map \#1, $w=1\meter$
and $\sigma=3\meter$ are assumed. The BCRB (\ref{eq:map1_bcrb}),
EZZB (\ref{eq:map1_zzb}) and WWB (\ref{eq:map1_wwb}) are shown for
comparison.\label{fig:map1_vs_deltaX}}
\end{figure}

\textbf{Bounds - }The BCRB, the EZZB and WWB for map \#1 are (see
(\ref{eq:bcrb_generic}), (\ref{eq:zzb_generic}), and (\ref{eq:wwb_generic}),
respectively) 
\begin{equation}
E_{x}\geq B_{x}=\sigma^{2}\frac{\rho^{2}}{\rho^{2}+\mathrm{J}_{s}}\label{eq:map1_bcrb}
\end{equation}
\begin{equation}
E_{x}\geq Z_{x}=\frac{\sigma^{2}}{4\rho}\left[2\zeta(\rho)+\zeta_{ov}(\rho_{\triangle},\rho,\rho)\right]\label{eq:map1_zzb}
\end{equation}
and $E_{x}\geq W_{x}$, respectively, where 
\begin{align}
W_{x} & =\sup_{h\in\mathbb{R}}\frac{h^{2}\exp\left(-\frac{h^{2}}{4\sigma^{2}}\right)\lambda^{2}\left(h,w,\triangle x\right)}{4w\left[\lambda\left(h,w,\triangle x\right)-\exp\left(-\frac{h^{2}}{2\sigma^{2}}\right)\gamma\left(h,w,\triangle x\right)\right]}\label{eq:map1_wwb}
\end{align}
$\rho\triangleq w/\sigma$ is the SNR, $\rho_{\triangle}\triangleq\triangle x/\sigma$,
\[
\lambda\left(h,w,\triangle x\right)\triangleq2\omega\left(w,h\right)+\omega_{ov}\left(\triangle x,w,w,h\right)
\]
and $\gamma\left(h,w,\triangle x\right)\triangleq2\omega\left(\frac{1}{2}w,h\right)$.
It is important to point out that that no approximations have been
adopted in the derivations of the BCRB, the EZZB and the WWB for the
considered 1-D scenario, so that (\ref{eq:map1_bcrb})-(\ref{eq:map1_wwb})
represent \emph{exact} bounds (of course, a smoothed uniform map has
been assumed for the BCRB only).\\
 \textbf{Estimators - }Map-aware localization can be accomplished
exploiting the MMSE estimator 
\begin{equation}
\hat{x}_{\text{\tiny{MMSE}}}(z)=(1/c(z))\sum_{n=1}^{N_{r}}\int_{\mathcal{R}_{n}}x\,\mathcal{N}\left(z;x,\sigma^{2}\right)dx\label{eq:MMSE_estimator_map_1}
\end{equation}
where $c(z)\triangleq\sum_{n=1}^{N_{r}}\int_{\mathcal{R}_{n}}\mathcal{N}\left(z;x,\sigma^{2}\right)dx$
is a normalization factor, or the maximum-a-posteriori (MAP) estimator
\begin{equation}
\hat{x}_{\text{\tiny{MAP}}}(z)=\arg\max_{\tilde{x}\in\mathcal{R}}\mathcal{N}\left(z;\tilde{x},\sigma^{2}\right)\label{eq:MAP_estimator_map_1}
\end{equation}
On the contrary, if the map is unknown, the (map-unaware) maximum
likelihood (ML) estimator $\hat{x}_{\text{\tiny{ML}}}(z)=z$ can be
used (obviously, the resulting estimate has the same variance $\sigma^{2}$
as the observation noise). Note that the RMSE performance of these
estimators is influenced by the geometrical parameters $w$ and $\triangle x$
and by the noise standard deviation $\sigma$. \\
 \textbf{Numerical results} \textbf{- }Figs. \ref{fig:map1_vs_noise}
and \ref{fig:map1_vs_deltaX} compare the square roots of the BCRB,
EZZB and WWB bounds $\sqrt{B_{x}}$, $\sqrt{Z_{x}}$, $\sqrt{W_{x}}$
and the RMSE of $\hat{x}_{\text{\tiny{MMSE}}}(z)$, $\hat{x}_{\text{\tiny{MAP}}}(z)$
and $\hat{x}_{\text{\tiny{ML}}}(z)$ versus $\rho$ (with $w$ and
$\triangle x$ fixed) and $\triangle x$ (with $w$ and $\sigma$
fixed), respectively. 

These results evidence that: 
\begin{enumerate}
\item The BCRB is tight at high SNRs only; in such conditions all the bounds
and the RMSE of both map-aware and map-unaware estimators tend to
$\sigma$ (see Fig. \ref{fig:map1_vs_noise}): conditional knowledge
dominates.
\item Unlike the BCRB, the EZZB and, in particular, the WWB exhibit a dependence
on $\triangle x$ and are much tighter (see Fig. \ref{fig:map1_vs_deltaX})
to the performance of optimal estimators and, for this reason, represent
more useful tools for predicting system accuracy.
\item The role played by map knowledge in estimation accuracy becomes significant
as $\rho$ decreases, i.e., as the quality of the observations gets
worse; indeed map knowledge prevents the bounds and the RMSE of the
map-aware estimators from diverging (note that in Fig. \ref{fig:map1_vs_noise}
the map-unaware RMSE diverges for $\rho\rightarrow0$). 
\item The performance gap among the MMSE, the MAP and especially the ML
estimators is significant for low SNRs (see Fig. \ref{fig:map1_vs_noise}). 
\item If the gap $\triangle x$ is small (i.e., if the segments $\mathcal{R}_{1}$
and $\mathcal{R}_{2}$ are close), the RMSE of the considered estimators
is low. On the contrary, if $\triangle x$ increases, the RMSE gets
larger, reaches a maximum and then decreases, tending to $w/\sqrt{12}$
(see Fig. \ref{fig:map1_vs_deltaX}). This behaviour can be explained
as follows: if the noise level is large with respect to $\triangle x$,
the estimate of the agent position may belong to the wrong segment,
thus resulting in a large error. However, a further increase of $\triangle x$
entails a reduction of the RMSE, since, for a large spacing between
$\mathcal{R}_{1}$ and $\mathcal{R}_{2}$, it is unlikely that the
wrong segment is selected in map-aware estimation. 
\end{enumerate}
\begin{figure}
\centering{}\includegraphics[width=3in]{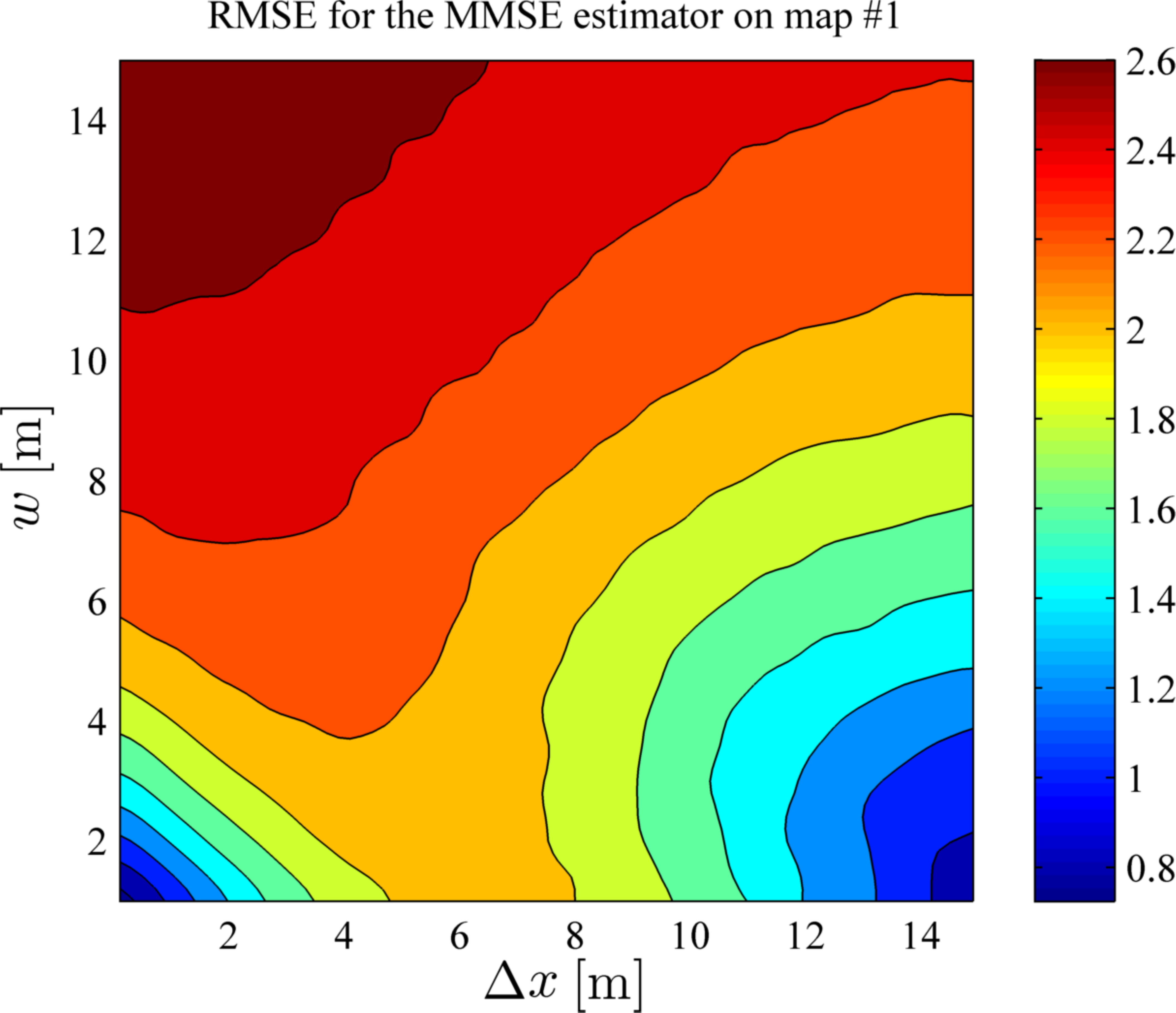}\caption{RMSE versus $\triangle x$ and $w$ for the MMSE estimator (\ref{eq:MMSE_estimator_map_1}).
Map \#1 and $\sigma=3\meter$ are assumed. The data has been smoothed
to remove simulation noise and improve readability.\label{fig:mmse_map1_surface}}
\end{figure}

Further insights are provided by Fig. \ref{fig:mmse_map1_surface},
which shows the RMSE versus $w$ and $\triangle x$ for the MMSE estimator
(note that the MMSE data shown in Fig. \ref{fig:map1_vs_deltaX} can
be extracted from Fig. \ref{fig:mmse_map1_surface} setting $w=1\meter$).
In fact, the numerical results shown in this figure evidence that: 
\begin{enumerate}
\item For $w>3\sigma$ the RMSE decreases only mildly as $\triangle x$
gets larger and the main geometric feature of the map affecting estimation
accuracy is the area of $\mathcal{R}$; in this case a map-aware estimator
still provides an advantage over map-unaware counterparts (the maximum
RMSE for the first estimator is $2.6\meter$, as shown in Fig. \ref{fig:mmse_map1_surface},
and is smaller than $\sigma=3\meter$, which represents the RMSE for
the ML estimator, as shown in Fig. \ref{fig:map1_vs_deltaX}). 
\item For $w<3\sigma$ the RMSE reaches its maximum for $\triangle x\simeq3\sigma-w$;
for this reason, any combination of the geometric parameters and noise
level satisfying such equality should be avoided. 
\end{enumerate}
Based on the numerical results shown in Fig. \ref{fig:mmse_map1_surface},
we can draw some observations about an indoor localization system
operating in two adjacent rooms, each having width $w_{room}$ and
separated by a wall having thickness $\triangle x_{wall}=0.2\meter$,
and processing noisy observations characterized by $\sigma=3\meter$
(realistic value for RSS systems, see \cite{Mazuelas2009b,Pivato2011,Liu2007}).
From Fig. \ref{fig:mmse_map1_surface} we can infer that, even if
$w_{room}<3\sigma=9\meter$, since $\triangle x_{wall}$ is appreciably
smaller than $3\sigma-w$, the advantage in terms of RMSE provided
by a map modelling wall obstructions is small with respect to that
offered by a map modelling the environment as a single connected room
having width $2w_{room}+\triangle x_{wall}$. In fact, in Fig. \ref{fig:mmse_map1_surface},
if the presence of the wall is accounted for, the map-aware estimator
performance is associated with the point $(\triangle x_{wall},w_{room})$,
whereas, if it is not, with the point $(0,w_{room}+\frac{1}{2}\triangle x_{wall})$.
Unless the room size $w_{room}$ is very small (i.e., $w_{room}\simeq\triangle x_{wall}$)
the two ``operating points'' will be very close (due to the small
value of $\triangle x_{wall}$) and the resulting RMSE values will
be similar so that, in this case, a map modelling wall obstructions
is not very useful. Note also that, if the quality of observations
improves substantially (i.e., $\sigma\rightarrow0$), map information
plays a less significant role than the conditional information provided
by such observations, so that a detailed a priori map modelling walls
is not useful in this case, too. On the contrary, maps may play an
important role in localization in outdoor environments, where the
region characterized by $w<3\sigma$ and $\triangle x\ll3\sigma-w$
in the $(\triangle x,w)$ plane is easier to reach, since noisy observations
and large obstructions should be expected; for this reason, in this
case the presence of the obstructions in map modelling should not
be neglected.

\subsection{Map \#2\label{sub:num_res_map2}}

\begin{figure*}[!t]

\normalsize
\setcounter{MYtempeqncnt}{\value{equation}}
\setcounter{equation}{33}

\begin{align}
\mathbf{E} & \succeq\mathbf{W}=\diag\left\{ \sup_{h_{x}\in\mathbb{R}}\frac{h_{x}^{2}\exp\left(-\frac{h_{x}^{2}}{4\sigma^{2}}\right)\lambda_{x}^{2}\left(h_{x},\mathcal{R}\right)}{2\mathcal{A_{R}}\left[\lambda_{x}\left(h_{x},\mathcal{R}\right)-\exp\left(-\frac{h_{x}^{2}}{2\sigma^{2}}\right)\gamma_{x}\left(h_{x},\mathcal{R}\right)\right]},\right.\left.\sup_{h_{y}\in\mathbb{R}}\frac{h_{y}^{2}\exp\left(-\frac{h_{y}^{2}}{4\sigma^{2}}\right)\lambda_{y}^{2}\left(h_{y},\mathcal{R}\right)}{2\mathcal{A_{R}}\left[\lambda_{y}\left(h_{y},\mathcal{R}\right)-\exp\left(-\frac{h_{y}^{2}}{2\sigma^{2}}\right)\gamma_{y}\left(h_{y},\mathcal{R}\right)\right]}\right\} \label{eq:map2_wwb}
\end{align}
\setcounter{equation}{\value{MYtempeqncnt}}
\hrulefill
\vspace*{4pt}
\end{figure*}

\begin{figure}
\centering{}\includegraphics[width=3.5in]{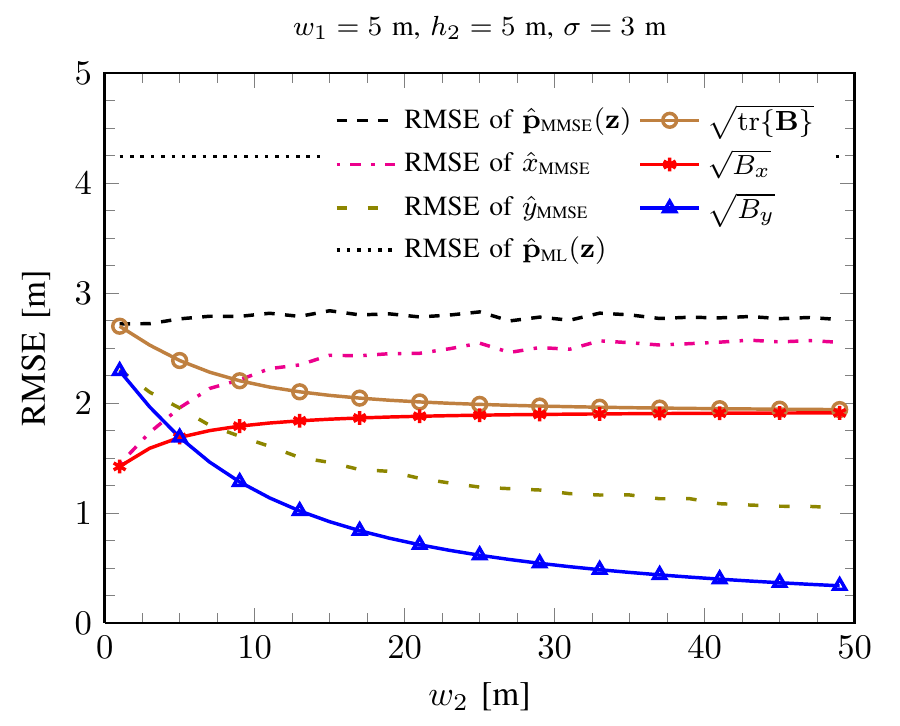}\caption{RMSE versus $w_{2}$ for the MMSE estimator (\ref{eq:MMSE_estimator_map_2}).
Map \#2 (Fig. \ref{fig:map_2d_lshape}), $\sigma=3\meter$ and $\mathcal{A_{R}}=75\meter^{2}$
are assumed. The BCRB components $\sqrt{B_{x}}$, $\sqrt{B_{y}}$,
$\sqrt{\trace\{\mathbf{B}\}}$ (see (\ref{eq:map2_bcrb})) and the
RMSE for the ML estimator are also shown for comparison.\label{fig:mmse_map2}}
\end{figure}

\textbf{Bounds - }Let us now focus on map \#2. The BCRB and the EZZB
in this case are (see (\ref{eq:bcrb_generic}) and (\ref{eq:zzb_generic}),
respectively) 
\begin{align}
\mathbf{E}\succeq\mathbf{B}=\diag & \left\{ \sigma^{2}\frac{\rho_{\mathcal{A}}^{2}}{\rho_{\mathcal{A}}^{2}+\mathrm{J}_{s}\left(\frac{h_{2}}{w_{1}}+\frac{h_{1}}{w_{1}+w_{2}}\right)},\right.\notag\\
 & \quad\left.\sigma^{2}\frac{\rho_{\mathcal{A}}^{2}}{\rho_{\mathcal{A}}^{2}+\mathrm{J}_{s}\left(\frac{w_{2}}{h_{1}}+\frac{w_{1}}{h_{1}+h_{2}}\right)}\right\} \label{eq:map2_bcrb}
\end{align}
\begin{align}
\mathbf{E}\succeq\mathbf{Z}=\diag & \left\{ \frac{\sigma^{2}}{2\rho_{\mathcal{A}}}\left[h_{1}\zeta\left(\frac{w_{1}+w_{2}}{\sigma}\right)+h_{2}\zeta\left(\frac{w_{1}}{\sigma}\right)\right],\right.\notag\\
 & \quad\left.\frac{\sigma^{2}}{2\rho_{\mathcal{A}}}\left[w_{1}\zeta\left(\frac{h_{1}+h_{2}}{\sigma}\right)+w_{2}\zeta\left(\frac{h_{1}}{\sigma}\right)\right]\right\} \label{eq:map2_zzb}
\end{align}
respectively, where $\rho_{\mathcal{A}}\triangleq\mathcal{A_{R}}/\sigma$.
The WWB derived from (\ref{eq:wwb_generic}) is shown in (\ref{eq:map2_wwb})
at the top of the next page
\addtocounter{equation}{1}; in (\ref{eq:map2_wwb}) the $\lambda_{x}\left(h_{x},\mathcal{R}\right)$
and $\gamma_{x}\left(h_{x},\mathcal{R}\right)$ functions have the
expressions 
\[
\lambda_{x}\left(h_{x},\mathcal{R}\right)=h_{1}\omega(w_{1}+w_{2},h_{x})+h_{2}\omega(w_{1},h_{x})
\]
\[
\gamma_{x}\left(h_{x},\mathcal{R}\right)=2h_{1}\omega\left(\frac{w_{1}+w_{2}}{2},h_{x}\right)+2h_{2}\omega\left(\frac{w_{1}}{2},h_{x}\right)
\]
respectively (dual expressions hold for $\lambda_{y}\left(h_{y},\mathcal{R}\right)$
and $\gamma_{y}\left(h_{y},\mathcal{R}\right)$). Note that $h_{1}$
and $h_{2}$ ($w_{1}$ and $w_{2}$) play a symmetric role in (\ref{eq:map2_bcrb}),
(\ref{eq:map2_zzb}) and (\ref{eq:map2_wwb}); this suggests that
a change in the geometrical features improving $B_{x}$ will reduce
$B_{y}$ and vice versa, when the overall area $\mathcal{A_{R}}$
of the map is kept constant.\\
 \textbf{Estimators - }Map-aware localization can be accomplished
exploiting the MMSE estimator 
\begin{equation}
\hat{\mathbf{p}}_{\text{\tiny{MMSE}}}(\mathbf{z})\triangleq\left[\begin{array}{c}
\hat{x}_{\text{\tiny{MMSE}}}\\
\hat{y}_{\text{\tiny{MMSE}}}
\end{array}\right]=(1/c(\mathbf{z}))\iint_{\mathcal{R}}\mathbf{p}\,\mathcal{N}\left(\mathbf{z};\mathbf{p},\boldsymbol{\Sigma}\right)d\mathbf{p}\label{eq:MMSE_estimator_map_2}
\end{equation}
where $c(\mathbf{z})\triangleq\iint_{\mathcal{R}}\mathcal{N}\left(\mathbf{z};\mathbf{p},\boldsymbol{\Sigma}\right)d\mathbf{p}$.
On the contrary, the (map-unaware) ML estimate is given by $\hat{\mathbf{p}}_{\text{\tiny{ML}}}(\mathbf{z})=\mathbf{z}$.\\
 \textbf{Numerical results - }Fig. \ref{fig:mmse_map2} shows the
square roots of the BCRB components $\sqrt{B_{x}}$, $\sqrt{B_{y}}$,
its trace $\sqrt{\trace\{\mathbf{B}\}}$ and the RMSE of the estimators
$\hat{x}_{\text{\tiny{MMSE}}}$, $\hat{y}_{\text{\tiny{MMSE}}}$,
$\hat{\mathbf{p}}_{\text{\tiny{MMSE}}}(\mathbf{z})$ and $\hat{\mathbf{p}}_{\text{\tiny{ML}}}(\mathbf{z})$
(the EZZB (\ref{eq:map2_zzb}) and WWB (\ref{eq:map2_wwb}) are not
shown to ease the reading); $\sigma=3\meter$, $w_{1}=5\meter$ and
$h_{2}=5\meter$ have been assumed in this case. In addition, when
increasing $w_{2}$, $h_{1}$ is reduced according to the law 
\[
h_{1}(w_{2})=\frac{\mathcal{A_{R}}-h_{2}w_{1}}{w_{1}+w_{2}}\,,
\]
so that the equality $\mathcal{A_{R}}=75\meter^{2}$ always holds.
From these results it is inferred that: 
\begin{enumerate}
\item As already noted for map \#1, map-unaware estimation, which is unaffected
by a change in the geometrical features of the map, provides a worse
accuracy than its map-aware counterpart (the RMSE of ML estimator
is $\simeq21\%$ larger than that of its MMSE counterpart).
\item Increasing $w_{2}$ and decreasing $h_{1}$ changes the aspect ratio
of a portion of the map (see Fig. \ref{fig:map_2d_lshape}); this
results in an appreciable reduction in the RMSE of the parameter $y$,
but also in a mild increase of the RMSE of the parameter $x$. Therefore,
the improvement along one direction is compensated for by the worsening
in the orthogonal direction, so that no significant advantage for
the RMSE of $\hat{\mathbf{p}}_{\text{\tiny{MMSE}}}(\mathbf{z})$ is
found. 
\end{enumerate}
The last result suggests that when the accuracy in the estimation
of a given element of $\mathbf{p}$ is improved maintaining $\mathcal{A_{R}}$
constant, a reduction in the estimation accuracy of the other element
should be expected, so that that the overall accuracy does not change
much. 

\subsection{Map \#3\label{sub:num_res_map3}}

\textbf{Observation model} \textbf{- }Finally, let us analyse the
RMSE performance of localization systems operating in the environment
described by map \#3 in two cases: A) the map-aware estimator knows
map \#3 exactly; B) the map-aware estimator assumes that the localization
environment is described by the bounding box of map \#3. Note that
the last case is significant from a practical point of view, since,
in the absence of detailed information about the floor of a building,
the bounding box represents the best a priori model which can be adopted
to describe the floor itself. \\
In addition, in this Paragraph model (\ref{eq:obs_model_gaussian})
is compared against a more specific counterpart, called ``ranging
observation model'' in the following; for such modelling we assume
that 1) the observations are acquired by 4 anchors placed in the map
corners; 2) the observation provided by the $i$-th anchor can be
expressed as%
\footnote{This model is commonly used in the study of localization systems exploiting
TOA and, with some slight changes, RSS.%
} 
\begin{equation}
z_{i}=\left\Vert \mathbf{p}-\mathbf{p}_{i}^{a}\right\Vert +n_{i}\label{eq:obs_model_ranging}
\end{equation}
with $i=1,2,3,4$, where $n_{i}$ represents Gaussian noise with variance
$\sigma^{2}$ and $\mathbf{p}_{i}^{a}$ represents the $i$-th anchor
position.

\begin{figure}
\centering{}\includegraphics[width=3.5in]{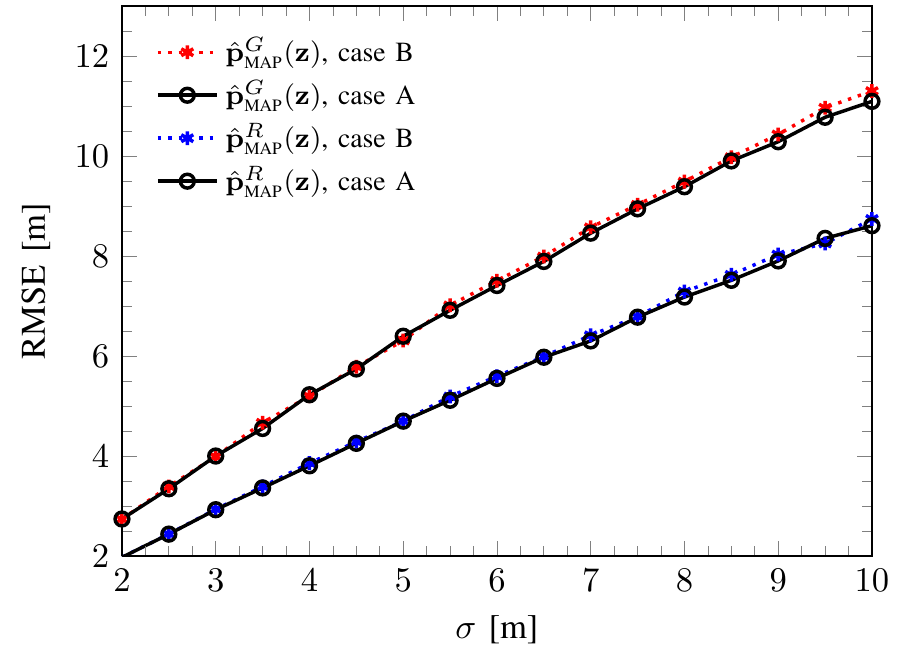}\caption{RMSE versus $\sigma$ for the MAP estimators (\ref{eq:MAP_estimator_map_3_toa})
and (\ref{eq:MAP_estimator_map_3_gaussian}), using either the Gaussian
observation model (see (\ref{eq:obs_model_gaussian})) or the ranging
observation model (see (\ref{eq:obs_model_ranging})). The a priori
map is map \#3 (case A) or its bounding box (case B). The RMSE of
ML estimators is also shown for comparison. \label{fig:mmse_map3}}
\end{figure}

The scenario considered in this example is helpful to assess 1) to
what extent a detailed knowledge of the geometrical features of a
map can improve localization accuracy with respect to the availability
of a simplified map model (like a bounding box model) and 2) if a
localization system based on the Gaussian model (\ref{eq:obs_model_gaussian})
can exhibit a similar behaviour as that of a localization system based
on (\ref{eq:obs_model_ranging}), at least in the considered cases.\textbf{}\\
\textbf{Estimators} \textbf{- }Given the model (\ref{eq:obs_model_ranging}),
the MAP estimator can be expressed as 
\begin{equation}
\hat{\mathbf{p}}_{\text{\tiny{MAP}}}^{R}(\mathbf{z})=\arg\max_{\tilde{\mathbf{p}}\in\mathcal{R}}\mathcal{N}\left(\mathbf{z};\boldsymbol{\mu}(\tilde{\mathbf{p}}),\boldsymbol{\Sigma}^{R}\right)\label{eq:MAP_estimator_map_3_toa}
\end{equation}
where $\boldsymbol{\mu}(\mathbf{p})\triangleq\left[\left\Vert \mathbf{p}-\mathbf{p}_{1}^{a}\right\Vert ,...,\left\Vert \mathbf{p}-\mathbf{p}_{4}^{a}\right\Vert \right]^{T}$
and $\boldsymbol{\Sigma}^{R}=\sigma^{2}\mathbf{I}_{4}$. If the observation
model (\ref{eq:obs_model_gaussian}) is adopted in place of (\ref{eq:obs_model_ranging}),
the MAP estimator 
\begin{equation}
\hat{\mathbf{p}}_{\text{\tiny{MAP}}}^{G}(\mathbf{z})=\arg\max_{\tilde{\mathbf{p}}\in\mathcal{R}}\mathcal{N}\left(\mathbf{z};\tilde{\mathbf{p}},\boldsymbol{\Sigma}^{G}\right)\label{eq:MAP_estimator_map_3_gaussian}
\end{equation}
is obtained, where $\boldsymbol{\Sigma}^{G}=\boldsymbol{\Sigma}=\sigma^{2}\mathbf{I}_{2}$.
\\
 \textbf{Numerical results} \textbf{- }Fig. \ref{fig:mmse_map3} compares
the RMSE performance achieved by the MAP estimators (\ref{eq:MAP_estimator_map_3_toa})
and (\ref{eq:MAP_estimator_map_3_gaussian}) versus $\sigma$ in cases
A and B. These results show that: 
\begin{enumerate}
\item Surprisingly the selection of the map model (map \#3 in case A or
its bounding box in case B) does not affect the MMSE estimation performance,
whatever the model for the observations. This result can be motivated
as follows. For each value of $\sigma$, there is some portion of
the map where the combination of noise level and geometrical features
(see the analysis of map \#1) entails a large RMSE and where a detailed
map (map \#3) does not provide a substantial advantage over simpler
maps (e.g., the bounding box); since the Bayesian RMSE results from
an average of the agent position over the entire map, it is biased
by such portions of the map for each value of $\sigma$. Such an effect
influences also any global Bayesian bound, like the BCRB, the EZZB
and the WWB: their values result from an average over the entire parameter
space (in this case $\mathcal{R}$) and can be strongly influenced
by the parameter values (in this case, the values of $\mathbf{p}$)
characterized by the largest errors \cite{Routtenberg2012}. Note
also that local bounds (e.g., CRB), which model the parameters to
be estimated as deterministic variables, do not suffer from this problem,
but cannot exploit a priori information and represent performance
limits for unbiased estimators only \cite{Bell1997}. In our case
study a detailed knowledge of map \#3 (case A) provides, in various
subsets of $\mathcal{R}$, a substantial improvement with respect
to the knowledge of the bounding box (case B), but the effects of
such subsets are then ``averaged out'' when evaluating the Bayesian
RMSE, as shown in Fig. \ref{fig:mmse_map3}.
\item The adoption of the Gaussian observation model (\ref{eq:obs_model_gaussian})
entails a worse accuracy than that provided by the ranging observation
model (\ref{eq:obs_model_ranging}). This can be related to the placement
of the anchors and to their number; in fact, range observations coming
from anchors displaced with a constant angular spacing allow to average
out the measurement errors better than the positional observation
of the Gaussian model (\ref{eq:obs_model_gaussian}) when the same
noise variance $\sigma^{2}$ is assumed in both models%
\footnote{It is important to note that using the same number of observations
for both models may generate misleading results. In fact, on the one
hand the minimal number of measurements usually considered for the
ranging observation model is 3 (this allows non-ambiguous 2-D localization);
our choice of 4 anchors and 4 measurements is perhaps more realistic.
On the other hand, for the Gaussian observation model, 2 positional
observations (one about $x$ and one about $y$) are considered since
2-D localization is performed.%
}. If the offset between the curves referring to the Gaussian and the
ranging model is neglected, our results show that both curves exhibit
the same behaviour, when $\sigma$ is varied; this is an hint to the
fact that the insights obtained from the bounds derived for the model
(\ref{eq:obs_model_gaussian}) may be useful also in the analysis
of systems employing different observation models (like (\ref{eq:obs_model_ranging})).
\end{enumerate}
Finally, comment 1) suggests that, when assessing the improvement
in localization accuracy coming from map awareness, the evaluation
of the overall RMSE does not provide sufficient information about
the importance of a given map. In fact, if the effects of different
portions of the maps are analysed, it is found that some of them play
a more relevant role than other ones; in this perspective, the insights
and the intuition acquired from the results referring to map \#1 are
very useful, given that map \#1 can be thought of as a transversal
slice of a subset of any 2-D map. 

\section{Conclusion\label{sec:conc}}

In this manuscript the impact of map awareness on localization performance
has been investigated from a theoretical perspective, evaluating different
accuracy bounds. Such bounds provide some general indications about
the role and importance of this form of a priori information. Our
study has evidenced that, unluckily, the tighter is an accuracy bound,
the more complicated is its dependence on the geometric parameters
of maps. This has motivated the analysis of the developed bounds in
specific environments; in particular, three different maps have been
considered and, for each of them, the accuracy bounds have been evaluated
and compared with the performance offered by map-aware and map-unaware
estimators. Our analytical and numerical results have evidenced that:
a) map-aware estimation accuracy can be related to some features of
the map (e.g., its shape and area) even though, in general, the relation
is complicated; b) maps are really useful in the presence of some
combination of low SNRs and specific geometrical features of the map
(e.g., the size of obstructions); c) for some combinations of SNRs
and geometrical features, there is no need of refined maps since knowledge
of map details provides only a negligible performance gain; d) in
a given environment the EZZB and, in particular, the WWB are usually
much tighter than the BCRB; e) in the cases of maps containing many
features, the importance of map awareness is better captured by the
analysis of subsets of those map. 

Future improvements include the development of map-aware bounds for
specific observation models; in particular models including bias due
to the non-line of sight (NLOS) propagation may unveil important insights
about map-aware localization systems operating in NLOS conditions.

Finally, it is important to point out that all the bounds derived
in this manuscript can be also exploited for other estimation problems
(provided that the vector of random parameters to be estimated is
characterized by a uniform a priori pdf) and can be easily generalized
to the case of estimation of an $N$-dimensional parameter vector,
with $N>2$.

\appendices{\label{apd:proofs_loc}}

\section{Derivation of the BFIM and BCRB \label{apd:proofs_bfim} }

In this Appendix the derivation of (\ref{eq:bfim_map_generic_mult_intervals})
and (\ref{eq:bfim_maps_nogap}) is sketched. To begin, we consider
a real function $s(t)$ having the following properties: 
\begin{lyxlist}{AAA}
\item [{P.1)}] it is continuous and differentiable; 
\item [{P.2)}] $s(t)\geq0\:\forall t\in\mathbb{R}$; 
\item [{P.3)}] $\int s(t)dt=1$; 
\item [{P.4)}] $\int\frac{\partial s(t)}{\partial t}dt=0$; 
\item [{P.5)}] its support is the interval%
\footnote{The measurement unit for the support of $s(t)$ is the same as that
adopted for $\mathbf{p}$ and $\mathbf{z}$ (see Sec. \ref{sec:scenario}).%
} $\left[-1/2;1/2\right]$;
\item [{P.6)}] $s(0)=1$. 
\end{lyxlist}
These properties have the following implications: a) $s(t)$ is a
pdf function (see P.1-P.3); b) an a-priori FI $\mathrm{J}_{s}$ can
be computed for it (the regularity condition P.4 ensures the FI existence)
\cite{kay}; c) $s(\cdot)$ can be used to model bounded statistical
distributions (i.e., maps), eventually after some scaling and translation
(see P.5, P.6). As far as the last point is concerned, it is worth
mentioning that 
\[
\tilde{s}(t,a,b)\triangleq(1/b)\cdot s\left((t-a)/b\right)\:,
\]
with $b>0$, is still a pdf function and its FI is $\tilde{\mathrm{J}}_{s}(b)=\mathrm{J}_{s}/b^{2}$.
Any function $s(\cdot)$ sharing the properties P.1-P.6 represents
a ``smoothing function''; specific examples of smoothing functions
can be found in \cite{techreport_buildingmaps}.

Any smoothing function $s(t)$ can be used to define the analytical
model of a smoothed uniform map. To show how this can be done, we
start considering a unidimensional (1-D) scenario first, where the
agent position to estimate is the scalar $x$. The support $\mathcal{R}\subset\mathbb{R}$
of a 1-D uniform map can be always represented as the union of $N_{r}$
disjoint segments spaced by $N_{r}-1$ segments where the map pdf
$f(x)$ is equal to 0. Let the $N_{r}$ segments of the support $\mathcal{R}$
be indexed by the odd numbers of the set $\mathcal{N}^{o}=\{1,3,...,2N_{r}-1\}$
and let $c{}_{n}$ ($w{}_{n}$) denote the centre (width) of the $n$-th
segment, with $n\in\mathcal{N}^{o}$. Then the map pdf $f(x)$ associated
with this scenario can be expressed as 
\begin{equation}
f(x)=\frac{1}{\mathcal{W_{R}}}\sum_{n\in\mathcal{N}^{o}}s\left(\frac{x-c_{n}}{w_{n}}\right)=\frac{1}{\mathcal{W_{R}}}\sum_{n\in\mathcal{N}^{o}}w_{n}\tilde{s}(x,c_{n},w_{n})\label{eq:pdf_1d_multirect}
\end{equation}
where $\mathcal{W_{R}}\triangleq\sum_{n\in\mathcal{N}^{o}}w_{n}$.
The a priori FI associated with $f(x)$ is given by 
\begin{align}
\mathrm{J}_{x} & \triangleq\mathbb{E}_{x}\left\{ \left(\frac{\partial\ln f(x)}{\partial x}\right)^{2}\right\} \nonumber \\
 & =\frac{1}{\mathcal{\mathcal{W_{R}}}}\sum_{n\in\mathcal{N}^{o}}w_{n}\mathbb{E}_{x}\left\{ \left(\frac{\partial}{\partial x}\ln\tilde{s}(x,c_{n},w_{n})\right)^{2}\right\} \nonumber \\
 & =\frac{1}{\mathcal{\mathcal{W_{R}}}}\sum_{n\in\mathcal{N}^{o}}w_{n}\tilde{\mathrm{J}}_{s}(w_{n})=\frac{1}{\mathcal{\mathcal{W_{R}}}}\sum_{n\in\mathcal{N}^{o}}\frac{\mathrm{J}_{s}}{w_{n}}\label{eq:fi_1d_map}
\end{align}
Let us extend these results to 2-D maps. Exploiting some definitions
given in Section \ref{sec:scenario} and assuming that the smoothing
along $x$ and that along $y$ are independent, the pdf for an arbitrary
smoothed uniform 2-D map can be put in the form%
\footnote{It can be easily proved that this pdf satisfies regularity condition
$\mathbb{E}_{\mathbf{p}}\left\{ \frac{\partial\ln f(\mathbf{p})}{\partial\mathbf{p}}\right\} =\mathbf{0}$.%
} 
\begin{align}
f(\mathbf{p})=\frac{1}{\mathcal{\mathcal{A_{R}}}} & \sum_{n\in\mathcal{N}_{h}^{o}(y)}s\left(\frac{x-c_{x,n}(y)}{w_{n}(y)}\right)\cdot\nonumber \\
 & \sum_{m\in\mathcal{N}_{v}^{o}(x)}s\left(\frac{y-c_{y,m}(x)}{h_{m}(x)}\right)\label{eq:pdf_2d_map}
\end{align}
where $c_{x,n}(y)$ ($c_{y,m}(x)$) and $w_{n}(y)$ ($h_{m}(x)$)
denote the centre and the length, respectively of the $n$-th ($m$-th)
segment. Let us now compute the elements of the BFIM $\mathbf{J}_{\mathbf{p}}$
associated with this pdf (see ($\ref{eq:bfim_apriori}$)). The analysis
of (\ref{eq:pdf_2d_map}) can be connected to that of (\ref{eq:pdf_1d_multirect})
exploiting the formula (iterated expectation) 
\begin{align}
\left[\mathbf{J}_{\mathbf{p}}\right]_{1,1} & \triangleq\mathbb{E}_{\mathbf{p}}\left\{ \left(\frac{\partial\ln f(\mathbf{p})}{\partial x}\right)^{2}\right\} \nonumber \\
 & =\mathbb{E}_{y}\left\{ \mathbb{E}_{x|y}\left\{ \left(\frac{\partial\ln f(\mathbf{p})}{\partial x}\right)^{2}\right\} \right\} \label{eq:bfim_2d_generic_map_x}
\end{align}
and adopting the approximation 
\begin{align}
f(\mathbf{p}) & \simeq\frac{1}{\mathcal{\mathcal{A_{R}}}}\sum_{n\in\mathcal{N}_{h}^{o}(y)}s\left(\frac{x-c_{x,n}(y)}{w_{n}(y)}\right)\label{eq:pdf_2d_map_approx}\\
 & =\frac{\mathcal{W_{R}}(y)}{\mathcal{\mathcal{A_{R}}}}\frac{1}{\mathcal{\mathcal{W_{R}}}(y)}\sum_{n\in\mathcal{N}_{h}^{o}(y)}s\left(\frac{x-c_{x,n}(y)}{w_{n}(y)}\right)\nonumber 
\end{align}
where $\mathcal{\mathcal{W_{R}}}(y)\triangleq\sum_{n\in\mathcal{N}_{h}^{o}(y)}w_{n}(y)$.
It is worth noting that in (\ref{eq:pdf_2d_map_approx}) the smoothing
along the $y$ coordinate is ignored. This is a reasonable approximation
whenever the smoothing only modifies a narrow area of the edges of
$\mathcal{R}$ compared to the area of the flat regions of (\ref{eq:pdf_2d_map});
the selection of a proper smoothing function ensures that such a condition
holds (see Sec. \ref{sub:scenario_map} and \cite{techreport_buildingmaps}).\\
Thanks to (\ref{eq:pdf_2d_map_approx}), the evaluation of the inner
expectation in (\ref{eq:bfim_2d_generic_map_x}) is equivalent to
that of the FI referring to a 1-D map which consists of $N_{r}=N_{h}(y)$
segments having widths $\left\{ w_{n}(y)\right\} $ and centred around
the points $\left\{ c_{x,n}(y)\right\} $; for this reason, from (\ref{eq:pdf_1d_multirect})-($\ref{eq:fi_1d_map}$)
it is easily inferred that 
\begin{equation}
\mathbb{E}_{x|y}\left\{ \left(\frac{\partial\ln f(\mathbf{p})}{\partial x}\right)^{2}\right\} \simeq\frac{\mathcal{\mathcal{\mathcal{W_{R}}}}(y)}{\mathcal{A_{R}}}\frac{\mathrm{J}_{s}}{\mathcal{\mathcal{\mathcal{W_{R}}}}(y)}\sum_{n\in\mathcal{N}_{h}^{o}(y)}\frac{1}{w_{n}(y)}\label{eq:bfim_2d_generic_map_inner_x}
\end{equation}
Substituting the last result in the RHS of (\ref{eq:bfim_2d_generic_map_x})
produces 
\begin{align}
\left[\mathbf{J}_{\mathbf{p}}\right]_{1,1} & \simeq\frac{\mathrm{J}_{s}}{\mathcal{\mathcal{A_{R}}}}\int_{\mathcal{Y}}\sum_{n\in\mathcal{N}_{h}^{o}(y)}\frac{1}{w_{n}(y)}f(y)dy\nonumber \\
 & \simeq\frac{\mathrm{J}_{s}}{\mathcal{\mathcal{A_{R}}}}\int_{\mathcal{Y}}\sum_{n\in\mathcal{N}_{h}^{o}(y)}\frac{dy}{w_{n}(y)}\label{eq:bfim_2d_generic_map_xbis}
\end{align}
where $f(y)\triangleq\int f(\mathbf{p})dx$ is the pdf of $y$ only
and it has been ignored in the integral since the smoothing only affects
a small portion of the integration domain (see Section \ref{sec:scenario}).
A similar approach can be adopted in the evaluation of $\left[\mathbf{J}_{\mathbf{p}}\right]_{2,2}$
(ignoring, in this case, the smoothing along $x$); this leads to
\begin{equation}
\left[\mathbf{J}_{\mathbf{p}}\right]_{2,2}\simeq\frac{\mathrm{J}_{s}}{\mathcal{\mathcal{A_{R}}}}\int_{\mathcal{X}}\sum_{m\in\mathcal{N}_{v}^{o}(x)}\frac{1}{h_{m}(x)}dx\label{eq:bfim_2d_generic_map_y}
\end{equation}
Note also that the approximated results (\ref{eq:bfim_2d_generic_map_xbis})-(\ref{eq:bfim_2d_generic_map_y})
become exact if $\mathcal{R}$ is a rectangle having its sides parallel
to the reference axes, since in this case the smoothing along $x$
is truly independent from the smoothing along $y$. For the same reason
the cross terms $\left[\mathbf{J}_{\mathbf{p}}\right]_{2,1}$ and
$\left[\mathbf{J}_{\mathbf{p}}\right]_{1,2}$ are exactly equal to
zero for a rectangle; however since 2-D maps usually exhibit a weak
correlation between the variables $x$ and $y$, these terms can be
neglected; this leads to (\ref{eq:bfim_map_generic_mult_intervals}),
from which (\ref{eq:bfim_maps_nogap}) is easily obtained assuming
that $\mathcal{N}_{h}^{e}(y)=\emptyset$ and $\mathcal{N}_{v}^{e}(x)=\emptyset$
$\forall x,y$.

\section{Derivation of the EZZB\label{apd:proofs_zzb}}

In this Appendix the EZZB ($\ref{eq:zzb_def}$) is derived following
a procedure conceptually similar to the one provided in \cite[Sec. II.B]{Bell1997}.
We start from the identity (see \cite[p. 24]{Cinlar1975}) 
\begin{equation}
E_{\nu}=\frac{1}{2}\int_{0}^{\infty}\Pr\left\{ \left|\xi_{\nu}\right|\geq\frac{h}{2}\right\} h\: dh\label{eq:zzb_1}
\end{equation}
where $\xi_{\nu}\triangleq\left[\hat{\mathbf{p}}(\mathbf{z})-\mathbf{p}\right]_{\nu}$.
Following \cite{Bell1997}, the probability appearing in (\ref{eq:zzb_1})
can be put in the form 
\begin{align}
\Pr\left\{ \left|\xi_{\nu}\right|\geq\frac{h}{2}\right\}  & =\Pr\left\{ \xi_{\nu}>\frac{h}{2}\right\} +\Pr\left\{ \xi_{\nu}\leq-\frac{h}{2}\right\} \nonumber \\
 & =\int_{\mathbb{R}^{N}}\Pr\left\{ \xi_{\nu}>\frac{h}{2}|\mathbf{p}=\boldsymbol{\rho}_{0}\right\} f(\boldsymbol{\rho}_{0})d\boldsymbol{\rho}_{0}\nonumber \\
 & +\int_{\mathbb{R}^{N}}\Pr\left\{ \xi_{\nu}\leq-\frac{h}{2}|\mathbf{p}=\boldsymbol{\rho}_{1}\right\} f(\boldsymbol{\rho}_{1})d\boldsymbol{\rho}_{1}\label{eq:zzb_2}
\end{align}
where $f(\cdot)$ denotes the pdf of $\mathbf{p}$. Let $\boldsymbol{\rho}_{0}=\boldsymbol{\rho}$
and $\boldsymbol{\rho}_{1}=\boldsymbol{\rho}+\boldsymbol{\delta}(h)$,
where $\boldsymbol{\delta}(h)$ is some function of the integration
variable of (\ref{eq:zzb_1}). If the integration domain of (\ref{eq:zzb_2})
is constrained to be a subset $\mathcal{R}(h)$ of the map support
$\mathcal{R}$, such that $f(\boldsymbol{\rho})>0$ and $f(\boldsymbol{\rho}+\boldsymbol{\delta}(h))>0$,
then, dividing and multiplying the RHS of (\ref{eq:zzb_2}) by $f(\boldsymbol{\rho})+f(\boldsymbol{\rho}+\boldsymbol{\delta}(h))>0$,
produces the expression \cite[eqs. (22)-(30)]{Bell1997} 
\begin{multline}
\Pr\left\{ \left|\xi_{\nu}\right|\geq\frac{h}{2}\right\} \geq\\
\int_{\mathcal{R}(h)}\left[f(\boldsymbol{\rho})+f(\boldsymbol{\rho}+\boldsymbol{\delta}(h))\right]P_{\text{\tiny{min}}}^{\mathbf{z}}(\boldsymbol{\rho},\boldsymbol{\rho}+\boldsymbol{\delta}(h))d\boldsymbol{\rho}\label{eq:zzb_3}
\end{multline}
($P_{\text{\tiny{min}}}^{\mathbf{z}}(\boldsymbol{\rho},\boldsymbol{\rho}+\boldsymbol{\delta}(h))$
is defined in Sec. \ref{sub:zzb}) which holds for any $\boldsymbol{\delta}(h)$
satisfying the equality $\mathbf{a}^{T}\boldsymbol{\delta}(h)=h$,
where $\mathbf{a}\in\mathbb{R}^{2}$. Note that (\ref{eq:zzb_3})
generalizes \cite[eq. (30)]{Bell1997} to the case where the pdf $f(\cdot)$
of the parameter to estimate has a bounded support. Here $\mathbf{a}=\mathbf{e}_{\nu}$
is selected so that $\boldsymbol{\delta}(h)=h\mathbf{e}_{\nu}$. Then
substituting (\ref{eq:zzb_3}) in (\ref{eq:zzb_1}) produces ($\ref{eq:zzb_def}$),
where the integration domains of $h$ and $\boldsymbol{\rho}$ have
been merged in the set $\mathcal{P}_{\nu}$.

Like in the previous Appendix, let us evaluate now ($\ref{eq:zzb_def}$)
for a 1-D uniform map whose support $\mathcal{R}\subset\mathbb{R}$
consists of the union of $N_{r}$ disjoint segments. In the following
such segments are indexed by the odd numbers of the set $\mathcal{N}^{o}=\{1,3,...,2N_{r}-1\}$
and the lower (upper) limit of the segment $n\in\mathcal{N}^{o}$
is denoted $l_{n}\triangleq c_{n}-\frac{w_{n}}{2}$ ($u_{n}\triangleq c_{n}+\frac{w_{n}}{2}$),
where $c{}_{n}$ ($w{}_{n}$) represents the centre (length) of the
segment itself. Moreover, it is assumed, without any loss of generality,
that $c_{1}<c_{3}<...<c_{2N_{r}-1}$. In this 1-D scenario ($\ref{eq:zzb_def}$)
simplifies as 
\begin{equation}
Z\triangleq\frac{1}{2}\iint_{\mathcal{P}}\left[f(\tau)+f(\tau+h)\right]P_{\text{\tiny{min}}}^{z}(\tau,\tau+h)h\, d\tau\: dh\label{eq:zzb_1d_1}
\end{equation}
where $\boldsymbol{\rho}=\tau$, 
\[
\mathcal{P}=\left\{ (\tau,h):h\geq0\wedge f(\tau)>0\wedge f(\tau+h)>0\right\} 
\]
 and $P_{\text{\tiny{min}}}^{z}(\tau,\tau+h)$ represents the minimum
error probability of a binary detector which, on the basis of a noisy
datum $z$ and a likelihood ratio test, has to select one of the following
two hypotheses $H_{0}:\: x=\tau$ and $H_{1}:\: x=\tau+h$. Note that,
for any $(\tau,h)\in\mathcal{P}$, we have that $f(\tau)+f(\tau+h)=2/\mathcal{W_{\mathcal{R}}}$
(with $\mathcal{W_{\mathcal{R}}}\triangleq\sum_{n\in\mathcal{N}^{o}}w_{n}\vphantom{A_{A_{A_{A_{A}}}}}$),
because of the uniformity of the considered map. If the Gaussian observation
model (\ref{eq:obs_model_gaussian}) adopted to the 1-D case is assumed,
the average error probability of the above mentioned detector is easily
shown to be 
\[
P_{\text{\tiny{min}}}^{z}(\tau,\tau+h)=\frac{1}{2}\erfc\left(h/(2\sigma\sqrt{2})\right)
\]
 (where $\sigma$ denotes the standard deviation of the noise observation
model), which is independent of $\tau$; then, (\ref{eq:zzb_1d_1})
simplifies as 
\begin{align}
Z & =\frac{1}{2\mathcal{\mathcal{W_{\mathcal{R}}}}}\iint_{\mathcal{P}}h\erfc\left(\frac{h}{2\sigma\sqrt{2}}\right)d\tau\: dh\nonumber \\
 & =\frac{\sigma^{3}}{2\mathcal{\mathcal{W_{\mathcal{R}}}}}\iint_{\mathcal{Q}}u\erfc\left(\frac{u}{2\sqrt{2}}\right)dt\: du\label{eq:zzb_1d_3}
\end{align}
where $u\triangleq h/\sigma$, $t\triangleq\tau/\sigma$ and 
\[
\mathcal{Q}\triangleq\left\{ (t,u):u\geq0\wedge f(\sigma t)>0\wedge f(\sigma(t+u))>0\right\} 
\]
 can be shown to be 2-D domain consisting of the union of $N_{r}$
triangles and $N_{r}(N_{r}-1)/2$ parallelograms in the plane $(t,u)$,
as exemplified by Fig. \ref{fig:appendix}, which refers to the case
$N_{r}=3$. In particular, the contribution to the RHS of (\ref{eq:zzb_1d_3})
from the $i$-th triangle 
\[
\mathcal{Q}_{i}\triangleq\left\{ (t,u):\frac{l_{i}}{\sigma}\leq t\leq\frac{u_{i}}{\sigma}\wedge0\leq u\leq\frac{u_{i}}{\sigma}-t\right\} 
\]
can be shown to be $\sigma^{2}\zeta(\rho_{i})/(2\rho_{\mathcal{W}})$,
where $\rho_{i}\triangleq w_{i}/\sigma$, $\rho_{\mathcal{W}}\triangleq\mathcal{W_{\mathcal{R}}}/\sigma$,
the function $\zeta(\rho)$ is defined in (\ref{eq:f_rho_def}) and
$i\in\mathcal{N}^{o}$ (see \cite{techreport_buildingmaps} for more
details). As far as the parallelograms are concerned, we retain only
the contributions coming from $(N_{r}-1)$ parallelograms 
\[
\mathcal{Q}_{ov,i}\triangleq\left\{ (t,u):\frac{l_{i}}{\sigma}\leq t\leq\frac{u_{i}}{\sigma}\right.\left.\wedge\frac{l_{i}}{\sigma}-t\leq u\leq\frac{u_{i}}{\sigma}-t\right\} 
\]
with $i\in\mathcal{N}^{e}\triangleq\{2,4,...,2(N_{r}-1)\}$ (see Fig.
\ref{fig:appendix}); such contributions are given by $\sigma^{2}\zeta_{ov}(\rho_{\triangle_{i}},\rho_{i-1},\rho_{i+1})/(2\rho_{\mathcal{W}})$,
where $\rho_{\triangle_{i}}\triangleq\triangle x_{i}/\sigma$, $\triangle x_{i}\triangleq l_{i+1}-u_{i-1}$
and $\zeta_{ov}(\rho_{\triangle},\rho_{1},\rho_{2})$ is defined in
(\ref{eq:f_gap_def}) for $\rho_{1}>\rho_{2}$ (it can be shown that
$\zeta_{ov}(\rho_{\triangle},\rho_{1},\rho_{2})=\zeta_{ov}(\rho_{\triangle},\rho_{2},\rho_{1})$).
Note that the contributions to (\ref{eq:zzb_1d_3}) coming from the
discarded parallelograms are always positive since $u\erfc\left(\frac{u}{2\sqrt{2}}\right)\geq0$
for $u\geq0$. Then, substituting the above mentioned contributions
in (\ref{eq:zzb_1d_3}) produces the lower bound 
\begin{equation}
Z>\frac{\sigma^{2}}{2\rho_{\mathcal{W}}}\left[\sum_{n\in\mathcal{N}^{o}}\zeta\left(\rho_{n}\right)+\sum_{n\in\mathcal{N}^{e}}\zeta_{ov}\left(\rho_{\triangle_{n}},\rho_{n-1},\rho_{n+1}\right)\right]\label{eq:zzb_1d}
\end{equation}
for the EZZB. The approach developed for a 1-D uniform map can be
extended to a 2-D scenario keeping into account that a) the term $P_{\text{\tiny{min}}}^{\mathbf{z}}(\boldsymbol{\rho},\boldsymbol{\rho}+h\mathbf{e}_{\nu})$
of ($\ref{eq:zzb_def}$) takes on a similar form as $P_{\text{\tiny{min}}}^{z}(\tau,\tau+h)$
of (\ref{eq:zzb_1d_1}), b) the integral on $\boldsymbol{\rho}$ can
be decomposed in a couple of nested integrals, one over $\mathcal{X}$
and the other one over $\mathcal{Y}$, and c) eq. (\ref{eq:zzb_1d})
can be exploited for the inner integral. Further details can be found
in \cite{techreport_buildingmaps}.

\section{Derivation of the WWB\label{apd:proofs_wwb}}

In this Appendix we first derive the WWB for a $2$-D uniform map
characterized by a support $\mathcal{R}$, assuming the observation
model (\ref{eq:obs_model_gaussian}). Note that this derivation can
be easily generalised to a $N$-D scenario. To begin, we note that,
since $f(\mathbf{z}|\mathbf{p})=\mathcal{N}(\mathbf{z};\mathbf{p},\boldsymbol{\Sigma})$
(see (\ref{eq:obs_model_gaussian})), the likelihood ratio $L\left(\mathbf{z};\mathbf{p}_{1},\mathbf{p}_{2}\right)$
(\ref{eq:likelihood_ratio}) is given by $\mathcal{N}(\mathbf{z};\mathbf{p}_{1},\boldsymbol{\Sigma})/\mathcal{N}(\mathbf{z};\mathbf{p}_{2},\boldsymbol{\Sigma})$
$\forall\mathbf{p}_{1},\mathbf{p}_{2}\in\mathcal{R}$. Then, it is
not difficult to prove that the expectation appearing in the numerator
of (\ref{eq:wwb_def}) is given by
\begin{align}
 & \mathbb{E}_{\mathbf{z},\mathbf{p}}\left\{ L^{\frac{1}{2}}\left(\mathbf{z};\mathbf{p}+h_{\nu}\mathbf{e}_{\nu},\mathbf{p}\right)\right\} \nonumber \\
 & =\frac{1}{\mathcal{A_{R}}}\int_{\mathbb{R}^{2}}\int_{\mathcal{P}_{\nu}(h_{\nu})}\negthickspace\mathcal{N}^{\frac{1}{2}}\left(\mathbf{z};\boldsymbol{\rho}+h_{\nu}\mathbf{e}_{\nu},\boldsymbol{\Sigma}\right)\mathcal{N}^{\frac{1}{2}}\left(\mathbf{z};\boldsymbol{\rho},\boldsymbol{\Sigma}\right)d\boldsymbol{\rho}\: d\mathbf{z}\nonumber \\
 & =\frac{1}{\mathcal{A_{R}}}\exp\left(-\frac{h_{\nu}^{2}}{8\sigma_{\nu}^{2}}\right)\int_{\mathcal{P}_{\nu}(h_{\nu})}d\boldsymbol{\rho}\label{eq:wwb_num}
\end{align}
where $\mathcal{P}_{\nu}(h_{\nu})\triangleq\left\{ \boldsymbol{\rho}:f(\boldsymbol{\rho})>0\wedge f(\boldsymbol{\rho}+h_{\nu}\mathbf{e}_{\nu})>0\right\} $
is a slice of the integration domain $\mathcal{P}_{\nu}\subset\mathbb{R}^{3}$
involved in the evaluation of the EZZB \cite{techreport_buildingmaps}
and $\nu\in\{x,y\}$. The expectation appearing in the denominator
of (\ref{eq:wwb}) can be expressed as 
\begin{multline*}
\mathbb{E}_{\mathbf{z},\mathbf{p}}\left\{ \left[L^{\frac{1}{2}}\left(\mathbf{z};\mathbf{p}+h_{\nu}\mathbf{e}_{\nu},\mathbf{p}\right)-L^{\frac{1}{2}}\left(\mathbf{z};\mathbf{p}-h_{\nu}\mathbf{e}_{\nu},\mathbf{p}\right)\right]^{2}\right\} \\
\shoveleft=\mathbb{E}_{\mathbf{z},\mathbf{p}}\left\{ L\left(\mathbf{z};\mathbf{p}+h_{\nu}\mathbf{e}_{\nu},\mathbf{p}\right)\right\} +\mathbb{E}_{\mathbf{z},\mathbf{p}}\left\{ L\left(\mathbf{z};\mathbf{p}-h_{\nu}\mathbf{e}_{\nu},\mathbf{p}\right)\right\} \\
-2\cdot\mathbb{E}_{\mathbf{z},\mathbf{p}}\left\{ L^{\frac{1}{2}}\left(\mathbf{z};\mathbf{p}+h_{\nu}\mathbf{e}_{\nu},\mathbf{p}\right)L^{\frac{1}{2}}\left(\mathbf{z};\mathbf{p}-h_{\nu}\mathbf{e}_{\nu},\mathbf{p}\right)\right\} 
\end{multline*}
i.e., as the sum of three terms which are denoted $A$, $B$ and $C$,
respectively, in the following. It is important to note that 
\begin{align}
A & =\mathbb{E}_{\mathbf{z},\mathbf{p}}\left\{ L\left(\mathbf{z};\mathbf{p}+h_{\nu}\mathbf{e}_{\nu},\mathbf{p}\right)\right\} \nonumber \\
 & =\frac{1}{\mathcal{A_{R}}}\int_{\mathbb{R}^{2}}\int_{\mathcal{P}_{\nu}(h_{\nu})}\mathcal{N}(\mathbf{z};\boldsymbol{\rho}+h_{\nu}\mathbf{e}_{\nu},\boldsymbol{\Sigma})d\boldsymbol{\rho}d\mathbf{z}\nonumber \\
 & =\frac{1}{\mathcal{A_{R}}}\int_{\mathcal{P}_{\nu}(h_{\nu})}d\boldsymbol{\rho}=\frac{1}{\mathcal{A_{R}}}\lambda_{\nu}\left(h_{\nu},\mathcal{R}\right)\nonumber \\
 & =\mathbb{E}_{\mathbf{z},\mathbf{p}}\left\{ L\left(\mathbf{z};\mathbf{p}-h_{\nu}\mathbf{e}_{\nu},\mathbf{p}\right)\right\} \nonumber \\
 & =B\label{eq:wwb_A_B}
\end{align}
and 
\begin{align}
C & =\frac{1}{\mathcal{A_{R}}}\int_{\mathbb{R}^{2}}\int_{\tilde{\mathcal{P}}_{\nu}(h_{\nu})}\mathcal{N}(\mathbf{z};\boldsymbol{\rho}+h_{\nu}\mathbf{e}_{\nu},\boldsymbol{\Sigma})d\boldsymbol{\rho}d\mathbf{z}\nonumber \\
 & =\frac{1}{\mathcal{A_{R}}}\exp\left(-\frac{h_{\nu}^{2}}{2\sigma_{\nu}^{2}}\right)\int_{\tilde{\mathcal{P}}_{\nu}(h_{\nu})}d\boldsymbol{\rho}\nonumber \\
 & =\frac{1}{\mathcal{A_{R}}}\exp\left(-\frac{h_{\nu}^{2}}{2\sigma_{\nu}^{2}}\right)\gamma_{\nu}\left(h_{\nu},\mathcal{R}\right)\label{eq:wwb_C}
\end{align}
where 
\begin{align*}
\tilde{\mathcal{P}}_{\nu}(h_{\nu})\triangleq\left\{ \boldsymbol{\rho}:f(\boldsymbol{\rho})>0\right. & \wedge f(\boldsymbol{\rho}+h\mathbf{e}_{\nu})>0\\
 & \left.\wedge f(\boldsymbol{\rho}-h\mathbf{e}_{\nu})>0\right\} 
\end{align*}
and the functions $\lambda_{\nu}\left(h_{\nu},\mathcal{R}\right)$
and $\gamma_{\nu}\left(h_{\nu},\mathcal{R}\right)$ are defined in
(\ref{eq:lambda_def}) and (\ref{eq:gamma_def}). Then, substituting
(\ref{eq:wwb_num}), (\ref{eq:wwb_A_B}), (\ref{eq:wwb_C}) in (\ref{eq:wwb_def})
produces (\ref{eq:wwb_generic}).

\begin{figure}
\begin{centering}
\includegraphics[width=3.5in]{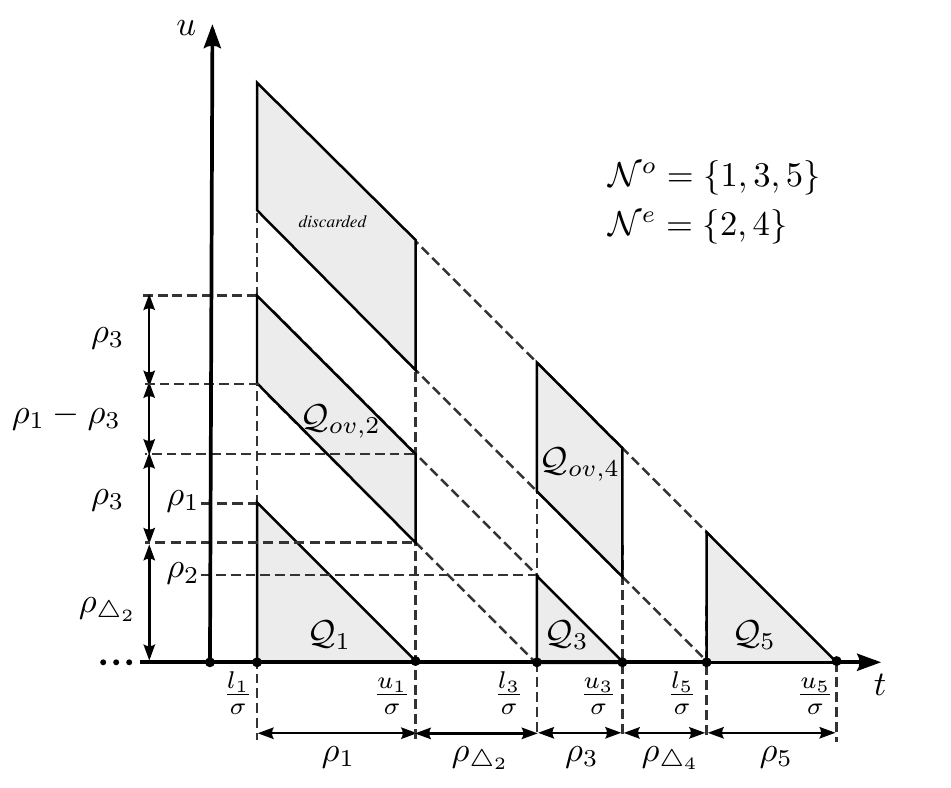}
\par\end{centering}

\caption{Representation of the set $\mathcal{Q}$ and of its subsets $\left\{ \mathcal{Q}_{i}\right\} _{i\in\mathcal{N}^{o}}$
and $\left\{ \mathcal{Q}_{ov,i}\right\} _{i\in\mathcal{N}^{e}}$ in
the $(t,u)$ plane ($N_{r}=3$ is assumed). Note that the contribution
from the ``discarded'' parallelogram is neglected in the evaluation
of both (\ref{eq:gamma_def})-(\ref{eq:lambda_def}) and (\ref{eq:zzb_1d_3}).
\label{fig:appendix}}
\end{figure}

Finally, it is worth pointing out that $\lambda_{\nu}\left(h_{\nu},\mathcal{R}\right)$
(\ref{eq:lambda_def}) and $\gamma_{\nu}\left(h_{\nu},\mathcal{R}\right)$
(\ref{eq:gamma_def}) can be explicitly related to the functions $\left\{ w_{n}(\cdot)\right\} $,
$\left\{ \triangle w_{n}(\cdot)\right\} $, $\left\{ h_{m}(\cdot)\right\} $,
$\left\{ \triangle h_{m}(\cdot)\right\} $ introduced in Section \ref{sec:scenario}.
To clarify this point, let us focus on $\lambda_{x}\left(h_{x},\mathcal{R}\right)$
and $\gamma_{x}\left(h_{x},\mathcal{R}\right)$ which can be derived,
for a given ordinate $y$, analysing the integration domain 
\[
\mathcal{P}_{x}(h_{x},y)\triangleq\left\{ \tau:f(\tau,y)>0\wedge f(\tau+h_{x},y)>0\right\} \:.
\]
If the variables $t\triangleq\tau/\sigma$ and $u\triangleq h_{x}/\sigma$
are defined, this domain can be put in the form 
\[
\mathcal{Q}_{x}(u,y)=\left\{ t:f(\sigma t,y)>0\wedge f(\sigma(t+u),y)>0\right\} \:,
\]
which describes a slice of the $(t,u)$ plane of Fig. \ref{fig:appendix}.
The contribution of the $i$-th triangle $\mathcal{Q}_{i}$ to $\lambda_{x}(h_{x},\mathcal{R}(y))$
($\gamma_{x}(h_{x},\mathcal{R}(y))$) is given by $\omega\left(w_{i}(y),h_{x}\right)$
($\omega_{ov}\left(\triangle w_{i}(y),w_{i-1}(y),w_{i+1}(y),h_{x}\right)$),
where $\omega(w_{n}(y),h_{x})$ is defined in (\ref{eq:omega_def}),
$\omega_{ov}\left(\triangle w,w_{1},w_{2},h\right)$ is defined in
(\ref{eq:omega_ov_def}) and $\mathcal{R}(y)$ is the slice of the
map support $\mathcal{R}$ at the ordinate $y$. Similarly, the contribution
from the $i$-th parallelogram $\mathcal{Q}_{ov,i}$ to $\lambda_{x}\left(h_{x},\mathcal{R}(y)\right)$
is given by $\omega_{ov}\left(\triangle w_{i}(y),w_{i-1}(y),w_{i+1}(y),h_{x}\right)$.
Therefore, the overall contribution to $\lambda_{x}(h_{x},\mathcal{R}(y))$
can be expressed as 
\begin{multline}
\sum_{n\in\mathcal{N}_{h}^{o}(y)}\omega\left(w_{n}(y),h_{x}\right)+\\
\sum_{n\in\mathcal{N}_{h}^{e}(y)}\omega_{ov}\left(\triangle w_{n}(y),w_{n-1}(y),w_{n+1}(y),h_{x}\right)\label{eq:sum1}
\end{multline}
whereas that to $\gamma_{x}(h_{x},\mathcal{R}(y))$ as 
\begin{equation}
2\sum_{n\in\mathcal{N}_{h}^{o}(y)}\omega\left(\frac{1}{2}w_{n}(y),h_{x}\right)\label{eq:sum2}
\end{equation}
Finally, integrating (\ref{eq:sum1}) and (\ref{eq:sum2}) produces
(\ref{eq:lambda_def}) and (\ref{eq:gamma_def}), respectively.

\vspace*{-0.5in}

\begin{IEEEbiographynophoto}
{Francesco Montorsi} (S’06) received both the Laurea degree (\emph{cum
laude}) and the Laurea Specialistica degree (\emph{cum laude}) in
Electronic Engineering from the University of Modena and Reggio Emilia,
Italy, in 2007 and 2009, respectively. He received the Ph.D degree
in Information and Communications Technologies (ICT) from the University
of Modena and Reggio Emilia in 2013.

He is employed as embedded system engineer for an ICT company since
2013. In 2011 he was a visiting PhD student at the Wireless Communications
and Network Science Laboratory of Massachusetts Institute of Technology
(MIT). His research interests are in the area of localization and
navigation systems, with emphasis on statistical signal processing
(linear and non-linear filtering, detection and estimation problems),
model-based design and model-based performance assessment.

Dr. Montorsi is a member of IEEE Communications Society and served
as a reviewer for the \textsc{IEEE Transactions on Wireless Communications},
\textsc{IEEE Transactions on Signal Processing}, \textsc{IEEE Wireless
Communications Letters} and several IEEE conferences.
\end{IEEEbiographynophoto}
\vspace*{-0.5in}

\begin{IEEEbiographynophoto}
{Santiago Mazuelas} (M'10) received his Ph.D. in mathematics and
Ph.D. in telecommunications engineering from the University of Valladolid,
Spain, in 2009 and 2011, respectively.

Since 2009, he has been a postdoctoral fellow in the Wireless Communication
and Network Sciences Laboratory at MIT. He previously worked from
2006 to 2009 as a researcher and project manager in a Spanish technological
center, as well as a junior lecturer in the University of Valladolid.
 His research interests are the application of mathematical and statistical
theories to communications, localization, and navigation networks.

Dr. Mazuelas has served as a member of the Technical Program Committee
(TPC) for the IEEE Global Communications Conference (GLOBECOM) in
2010, 2011, and 2012, the IEEE International Conference on Ultra-Wideband
(ICUWB) in 2011, the 75th IEEE Vehicular Technology Conference 2012,
and the IEEE Wireless Telecommunications Symposium (WTS) in 2010 and
2011. He has received the 2012 IEEE Communications Society Fred W.
Ellersick Prize, the Best Paper Award from the IEEE GLOBECOM in 2011,
 the Best Paper Award from the IEEE ICUWB in 2011, and the young scientists
prize for the best communication at the Union Radio-Scientifique Internationale
(URSI) XXII Symposium in 2007.\vspace{1.8in}
\newpage{}
\end{IEEEbiographynophoto}

\begin{IEEEbiographynophoto}
{Giorgio M. Vitetta} (S’89-M’91-SM’99) was born in Reggio Calabria,
Italy, in April 1966. He received the Dr. Ing. degree in electronic
engineering (\emph{cum laude}) in 1990, and the Ph.D. degree in 1994,
both from the University of Pisa, Pisa, Italy. From 1995 to 1998,
he was a Research Fellow with the Department of Information Engineering,
University of Pisa. From 1998 to 2001, he was an Associate Professor
with the University of Modena and Reggio Emilia, Modena, Italy, where
he is currently a full Professor. 

His main research interests lie in the broad area of communication
theory, with particular emphasis on detection/equalization/synchronization
algorithms for wireless communications, statistical modelling of wireless
and powerline channels, ultrawideband communication techniques and
applications of game theory to wireless communications. 

Dr. Vitetta is serving as an Editor of the \textsc{IEEE Wireless Communications
Letters} and as an Area Editor of the \textsc{IEEE Transactions on
Communications}.
\end{IEEEbiographynophoto}
\vspace*{-0.5in}

\begin{IEEEbiographynophoto}
{Moe Z. Win} 
%
%
(S'85-M'87-SM'97-F'04) 
received 
	both the Ph.D.\ in 
		\switchAtwo{Electrical Engineering}
				{Electrical Engineering (under the supervision of Professor Robert A.\ Scholtz)}
	and M.S.\ in 
		\switchAtwo{Applied Mathematics}
				{Applied Mathematics (under the supervision of Professor Solomon W.\ Golomb)}
		as a Presidential Fellow at the University of Southern California (USC) in 1998.
He received an M.S.\ in Electrical Engineering from USC in 1989, 
	and a B.S.\ ({\em magna cum laude}) in Electrical Engineering from Texas A\&M University in 1987.

%
%
He is a Professor at the Massachusetts Institute of Technology (MIT).
Prior to joining MIT, he was at AT\&T Research Laboratories for five years 
and at the Jet Propulsion Laboratory for seven years.
%
%
His research encompasses fundamental theories, algorithm design, and
experimentation for a broad range of real-world problems.
His current research topics include
	network localization and navigation, 
	network interference exploitation, 
	intrinsic wireless secrecy,
	adaptive diversity techniques,
		\switchBtwo{and ultra-wide bandwidth systems.}
				{ultra-wide bandwidth systems,
				optical transmission systems, 
				and 
				space communications systems.}
		
%
%
Professor Win is 
	an elected Fellow of the AAAS, the IEEE, and the IET, 
	and was an IEEE Distinguished Lecturer.
He was honored with two IEEE Technical Field Awards: 
	the IEEE Kiyo Tomiyasu Award (2011) 
		\switchBtwo{and}
				{for ``fundamental contributions to high-speed reliable communications over optical and wireless channels''  and}
	the IEEE Eric E. Sumner Award 
		\switchBtwo{(2006, jointly with R.\ A.\ Scholtz).}
				{(2006, jointly with R.\ A.\ Scholtz) for ``fundamental contributions to high-speed reliable communications over optical and wireless channels.''}
%
%
Together with students and colleagues, his papers have received numerous awards including
	\switchBtwo{the IEEE Communications Society's Stephen O.\ Rice Prize (2012),
			the IEEE Aerospace and Electronic Systems Society's M.\ Barry Carlton Award (2011),
			the IEEE Communications Society's Guglielmo Marconi Prize Paper Award (2008),
    			and the IEEE Antennas and Propagation Society's Sergei A.\ Schelkunoff Transactions Prize Paper Award (2003).}
			{the IEEE Communications Society's Stephen O.\ Rice Prize (2012),
			the IEEE Communications Society's William R.\ Bennett Prize (2012),
			the IEEE Communications Society's Fred W.\ Ellersick Prize (2012),
			the IEEE Communications Society's Leonard G.\ Abraham Prize (2011),
			the IEEE Aerospace and Electronic Systems Society's M.\ Barry Carlton Award (2011),
			the IEEE Communications Society's Guglielmo Marconi Prize Paper Award (2008),
    			and the IEEE Antennas and Propagation Society's Sergei A.\ Schelkunoff Transactions Prize Paper Award (2003).}
%
%
Highlights of his international scholarly initiatives are
	the Copernicus Fellowship (2011),
	the Royal Academy of Engineering Distinguished Visiting Fellowship (2009),
	and
	the Fulbright
		\switchBtwo{Fellowship (2004).}
				{Foundation Senior Scholar Lecturing and Research Fellowship (2004).}
%
%
Other recognitions include
	\switchBtwo{the Laurea Honoris Causa from the University of Ferrara (2008),	
				the Technical Recognition Award of the IEEE ComSoc Radio Communications Committee (2008),
        				and the U.S. Presidential Early Career Award for Scientists and Engineers (2004).}	
			{the Outstanding Service Award of the IEEE ComSoc Radio Communications Committee (2010),
				the Laurea Honoris Causa from the University of Ferrara, Italy (2008),
				the Technical Recognition Award of the IEEE ComSoc Radio Communications Committee (2008),
        				the Wireless Educator of the Year Award (2007),
				the U.S. Presidential Early Career Award for Scientists and Engineers (2004), 
				the AIAA Young Aerospace Engineer of the Year (2004),
				and
				the Office of Naval Research Young Investigator Award (2003).}

%
%
Dr.\ Win is an elected Member-at-Large on the IEEE Communications Society Board of Governors (2011--2013).
He was
    the chair (2004--2006) and secretary (2002--2004) for
        the Radio Communications Committee of the IEEE Communications Society.
%
%
\switchBtwo{Over the last decade, he has organized and chaired numerous international conferences.} 
		{He served as
			the Technical Program Chair for
				the IEEE Wireless Communications and Networking Conference (2009),
				the IEEE Conference on Ultra Wideband (2006),
				the IEEE Communication Theory Symposia of ICC (2004) and Globecom (2000),
				 and
				the IEEE Conference on Ultra Wideband Systems and Technologies (2002);
			Technical Program Vice-Chair for
				the IEEE International Conference on Communications (2002); and
			the Tutorial Chair for
				ICC (2009) and
				the IEEE Semiannual International Vehicular Technology Conference (Fall 2001).}
%
%
He is currently
	an Editor-at-Large for the 
	{\scshape IEEE Wireless Communications Letters},
	and serving on the Editorial Advisory Board for the 
	{\scshape IEEE Transactions on Wireless Communications}.
He served as Editor (2006--2012) for
	\switchBtwo{the {\scshape IEEE Transactions on Wireless Communications},
				and served as 
				Area Editor (2003--2006) and Editor (1998--2006)
				for the {\scshape IEEE Transactions on Communications}.}      	
			  {the {\scshape IEEE Transactions on Wireless Communications}.
				He also served as the
	    			Area Editor for {\em Modulation and Signal Design} (2003--2006),
	    			Editor for {\em Wideband Wireless and Diversity} (2003--2006), and
	    			Editor for {\em Equalization and Diversity} (1998--2003),
	        			all for the {\scshape IEEE Transactions on Communications}.}
He was Guest-Editor
        for the
        {\scshape Proceedings of the IEEE}
		\switchBtwo{(2009)}
				{(Special Issue on UWB Technology \& Emerging Applications -- 2009)} and
        {\scshape IEEE Journal on Selected Areas in Communications}
        		 \switchBtwo{(2002).}
		 		{(Special Issue on Ultra\thinspace-Wideband Radio in Multiaccess Wireless Communications  -- 2002).}

\vspace{1.5in}
\end{IEEEbiographynophoto}

\end{document}